\documentclass{aastex61}
\usepackage{subfigure}
\usepackage{xspace}
\usepackage{graphicx}
\usepackage{natbib}
\usepackage{xcolor, colortbl}
\usepackage{url}
\definecolor{gainsboro}{rgb}{0.86, 0.86, 0.86}
\usepackage{lineno}

\newcommand{\Fermi}{\emph{Fermi}\xspace}
\newcommand{\Swift}{\emph{Swift}\xspace}
\newcommand{\Tninety}{$\rm T_{90}$\xspace}
\newcommand{\Epeak}{$E_{\rm peak}$\xspace}

\newcommand{\mod}[1]{#1}
\begin{document}

\title{An Ordinary Short Gamma-Ray Burst with Extraordinary Implications: \\ \Fermi-GBM Detection of GRB~170817A}

\correspondingauthor{A.~Goldstein}
\email{Adam.M.Goldstein@nasa.gov}
\author{A.~Goldstein}
\affiliation{Science and Technology Institute, Universities Space Research Association, Huntsville, AL 35805, USA}

\author{P.~Veres}
\affiliation{Center for Space Plasma and Aeronomic Research, University of Alabama in Huntsville, 320 Sparkman Drive, Huntsville, AL 35899, USA}

\author{E.~Burns}
\affiliation{NASA Postdoctoral Program Fellow, Goddard Space Flight Center, Greenbelt, MD 20771, USA}

\author{M.~S.~Briggs}
\affiliation{Space Science Department, University of Alabama in Huntsville, 320 Sparkman Drive, Huntsville, AL 35899, USA}
\affiliation{Center for Space Plasma and Aeronomic Research, University of Alabama in Huntsville, 320 Sparkman Drive, Huntsville, AL 35899, USA}

\author{R.~Hamburg}
\affiliation{Space Science Department, University of Alabama in Huntsville, 320 Sparkman Drive, Huntsville, AL 35899, USA}
\affiliation{Center for Space Plasma and Aeronomic Research, University of Alabama in Huntsville, 320 Sparkman Drive, Huntsville, AL 35899, USA}

\author{D.~Kocevski}
\affiliation{Astrophysics Office, ST12, NASA/Marshall Space Flight Center, Huntsville, AL 35812, USA}

\author{C.~A.~Wilson-Hodge}
\affiliation{Astrophysics Office, ST12, NASA/Marshall Space Flight Center, Huntsville, AL 35812, USA}

\author{R.~D.~Preece}
\affiliation{Space Science Department, University of Alabama in Huntsville, 320 Sparkman Drive, Huntsville, AL 35899, USA}
\affiliation{Center for Space Plasma and Aeronomic Research, University of Alabama in Huntsville, 320 Sparkman Drive, Huntsville, AL 35899, USA}

\author{S.~Poolakkil}
\affiliation{Space Science Department, University of Alabama in Huntsville, 320 Sparkman Drive, Huntsville, AL 35899, USA}
\affiliation{Center for Space Plasma and Aeronomic Research, University of Alabama in Huntsville, 320 Sparkman Drive, Huntsville, AL 35899, USA}

\author{O.~J.~Roberts}
\affiliation{Science and Technology Institute, Universities Space Research Association, Huntsville,
AL 35805, USA}

\author{C.~M.~Hui}
\affiliation{Astrophysics Office, ST12, NASA/Marshall Space Flight Center, Huntsville, AL 35812, USA}

\author{V.~Connaughton}
\affiliation{Science and Technology Institute, Universities Space Research Association, Huntsville,
AL 35805, USA}

\author{J.~Racusin}
\affiliation{NASA Goddard Space Flight Center, Greenbelt, MD 20771, USA}

\author{A.~von~Kienlin}
\affiliation{Max-Planck-Institut f\"{u}r extraterrestrische Physik, Giessenbachstrasse 1, 85748 Garching, Germany}

\author{T. Dal Canton}
\affiliation{NASA Postdoctoral Program Fellow, Goddard Space Flight Center, Greenbelt, MD 20771, USA}

\author{N.~Christensen}
\affiliation{Physics and Astronomy, Carleton College, MN 55057, USA}
\affiliation{Artemis, Universit\'{e} C\^{o}te d'Azur, Observatoire C\^{o}te d'Azur, CNRS, CS 34229, F-06304 Nice Cedex 4, France}

\author{T.~Littenberg}
\affiliation{Astrophysics Office, ST12, NASA/Marshall Space Flight Center, Huntsville, AL 35812, USA}

\author{K. Siellez}
\affiliation{Center for Relativistic Astrophysics and School of Physics, Georgia Institute of Technology, Atlanta, GA 30332, USA}

\author{L.~Blackburn}
\affiliation{Harvard-Smithsonian Center for Astrophysics, 60 Garden St, Cambridge, MA 02138, USA}

\author{J.~Broida}
\affiliation{Physics and Astronomy, Carleton College, MN 55057, USA}

\author{E.~Bissaldi}
\affiliation{Istituto Nazionale di Fisica Nucleare, Sezione di Bari, I-70126 Bari, Italy}

\author{W.~H.~Cleveland}
\affiliation{Science and Technology Institute, Universities Space Research Association, Huntsville, AL 35805, USA}

\author{M.~H.~Gibby}
\affiliation{Jacobs Technology, Inc., Huntsville, AL 35805, USA}

\author{M.~M.~Giles}
\affiliation{Jacobs Technology, Inc., Huntsville, AL 35805, USA}

\author{R.~M.~Kippen}
\affiliation{Los Alamos National Laboratory, PO Box 1663, Los Alamos, NM 87545, USA}

\author{S.~McBreen}
\affiliation{School of Physics, University College Dublin, Belfield, Stillorgan Road, Dublin 4, Ireland}

\author{J.~McEnery}
\affiliation{NASA Goddard Space Flight Center, Greenbelt, MD 20771, USA}

\author{C.~A.~Meegan}
\affiliation{Center for Space Plasma and Aeronomic Research, University of Alabama in Huntsville, 320 Sparkman Drive, Huntsville, AL 35899, USA}

\author{W.~S.~Paciesas}
\affiliation{Science and Technology Institute, Universities Space Research Association, Huntsville, AL 35805, USA}

\author{M.~Stanbro}
\affiliation{Space Science Department, University of Alabama in Huntsville, 320 Sparkman Drive, Huntsville, AL 35899, USA}

\begin{abstract}
On August 17, 2017 at 12:41:06 UTC the \Fermi Gamma-ray Burst Monitor (GBM) detected and triggered on the short gamma-ray burst GRB~170817A. Approximately 1.7~s prior to this GRB, the Laser Interferometer Gravitational-Wave Observatory (LIGO) triggered on a binary compact merger candidate associated with the GRB. This is the first unambiguous coincident observation of gravitational waves and electromagnetic radiation from a single astrophysical source and marks the start of gravitational-wave multi-messenger astronomy. We report the GBM observations and analysis of this ordinary short GRB, which extraordinarily confirms that at least some short GRBs are produced by binary compact mergers.
\end{abstract}

 \section{Introduction}
Since beginning operations in July 2008, the \Fermi Gamma-ray Burst Monitor (GBM) has autonomously detected over 2000 gamma-ray bursts (GRBs), providing real-time alerts, degree-precision sky localizations, and high-quality data for temporal and spectral analysis. The wide field-of-view and high uptime of GBM make it a key instrument for detecting electromagnetic (EM) counterparts to gravitational wave (GW) signals, facilitating broad scientific analyses of coincident events.

\begin{figure}
	\includegraphics[width=15cm]{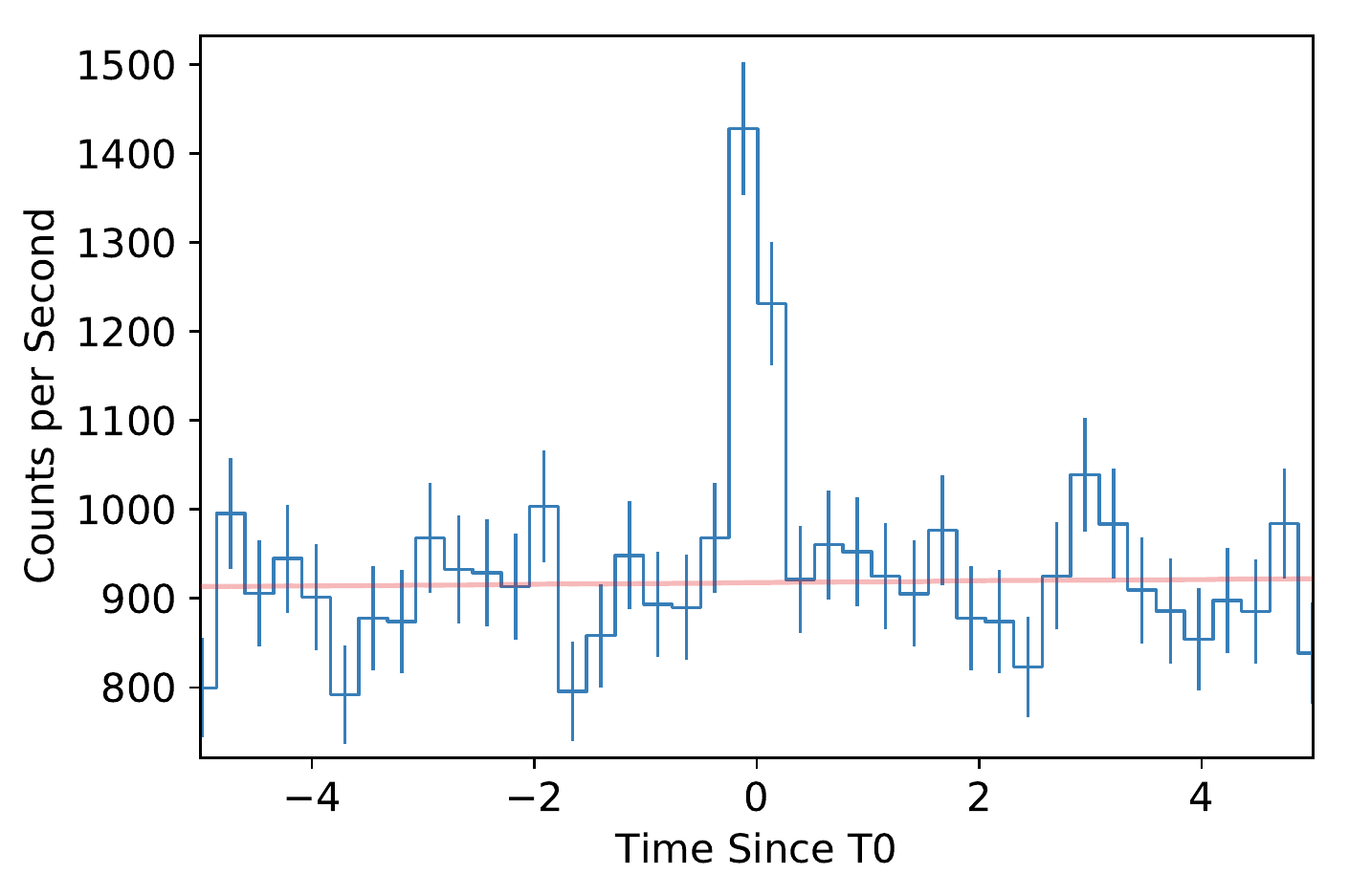}
	\caption{The 256 ms binned lightcurve of GRB~170817A in the 50--300 keV band for NaIs 1, 2, and 5. The red band is the un-binned Poisson maximum likelihood estimate of the background.}
	\label{fig:BTTE_sum_zoom}
\end{figure}

The GBM detection of GRB~170817A, shown in Figure~\ref{fig:BTTE_sum_zoom}, was not extraordinary. GBM detected the GRB in orbit in real time, a process referred to as ``triggering'', performed on-board classification and localization, and transmitted the results to EM and GW follow-up partners within seconds, as it has done for thousands of other transients. This particular trigger was different in one important aspect: a coincident GW trigger by the Laser Interferometer Gravitational Wave Observatory (LIGO) occurred $\sim$1.7 seconds prior to the GBM trigger~\citep{gcnLVCGW170817_1}, marking the first confident joint EM-GW observation in history.

The GW observation yielded a localization incorporating information from the two LIGO detectors, L1 \& H1, and the {\it Virgo} detector, V1, and is therefore termed an HLV map.  This initial HLV map was produced by the BAYESTAR algorithm~\citep{bayestar}, with a location centroid at RA = 12h57m, Dec = -17d51m and a 50\% (90\%) credible region spanning 9 (31) square degrees.  An estimate for the luminosity distance was also reported as $40 \pm 8$ Mpc~\citep{gcnLVCGW170817_2}.  An updated map incorporating Monte Carlo parameter estimation from the LALInference algorithm~\citep{veitch2015parameter} yielded a centroid at RA = 13h09m, Dec = -25d37m and a 50\% (90\%) credible region covering 8.6 (33.6) square degrees~\citep{gcnLVCGW170817_LAL}.  Another gamma-ray instrument, the Anti-Coincidence Shield for the SPectrometer for Integral (SPI-ACS) also detected GRB~170817A as a weak, $> 3\sigma$ S/N signal coincident in time to the GBM trigger~\citep{gcnACSGRB170817A_1,acs170817apaper}.  Utilizing the time difference between the GBM and SPI-ACS signals and the known positions of the parent spacecrafts, the Inter-Planetary Network~\citep[IPN;][]{IPN3} calculated an annulus on the sky which was consistent with both the GBM localization and the HLV map~\citep{gcnIPN}. Additionally, $\sim 12$ hours after the GBM and LIGO alerts, the discovery of a possible associated optical transient (OT) was reported~\citep{swope_sss17a,coulterinprep} and confirmed~\citep{decam_sss17a,decaminprep}, consistent with the location and distance reported by LIGO/{\it Virgo}.  The position of the OT is RA = 13h09m48.085s, Dec = -23d22m53.343s.  

We detail the GBM observations of this event in the following manner: description of the GBM instrument and its capabilities, discussion of the trigger and prompt localization of the GRB, and results of the standard analyses that are performed for every triggered GRB so that this GRB can be easily compared to the population which GBM observes. Due to the important nature of the joint detection, we proceed beyond the standard analyses and present more detailed analyses of the duration, pulse shape, spectrum, and searches for other associated gamma-ray emission.  We also determine how much dimmer the GRB could have been and still have been detected by GBM.  For these analyses  we assume the position of the OT.  Further use of the distance, redshift, and other information from the GW and EM-followup observations to perform rest-frame calculations are left to a companion analysis~\citep{jointpaper} which, in part, relies on the GBM analysis provided here.

\section{GBM Description}
GBM is one of two instruments on-board the \Fermi Gamma-Ray Space Telescope and is comprised of 14 detectors designed to study the gamma-ray sky in the energy band of $\sim$8 keV--40 MeV~\citep{2009Meegan}.  Twelve of the detectors are thalium-doped sodium iodide (NaI) scintillation detectors, which cover an energy range of 8--1000 keV and are pointed at various angles in order to survey the entire sky unocculted by the Earth at any time during the orbit. The relative signal amplitudes in the NaI detectors are also used to localize transients~\citep{2015ApJS..216...32C}.  The other two detectors are composed of bismuth germanate (BGO) crystals, cover an energy range of 200 keV--40 MeV, and are placed on opposite sides of the spacecraft. Incident photons interact with the NaI and BGO crystals and create scintillation photons. Those photons are then collected by the attached photomultiplier tubes and converted into electronic signals.  A recorded signal, which might be a gamma-ray or charged particle, is termed a count. 

Several data types are produced on-board GBM by binning the counts into pre-defined timescales (continuous/trigger): CTIME (256 ms/64 ms), CSPEC (4096/1024 ms), and TRIGDAT (variable, only by trigger; 64 ms--8.192 s).  During the first several years of the mission, data on individual counts, termed  Time-Tagged Events (TTE) were only produced during on-board triggers. An increase in telemetry volume and a flight software update in November 2012 allowed downlinking TTE data with continuous coverage for offline analysis.  The TTE data type is especially useful as it provides arrival time information for individual photons at 2 $\mu$s precision.  Additionally, while the CTIME and TRIGDAT data types only have a coarse energy resolution of 8 channels, the CSPEC and TTE data both have 128-channel energy resolution, facilitating spectral analysis of GRBs and other high-energy astrophysical, solar, and terrestrial phenomena.

\section{GBM Trigger \& Localization}
The flight software on-board GBM monitors the detector rates and triggers when a statistically significant rate increase occurs in two or more NaI detectors. Currently, 28 combinations of timescales and energy ranges are tested; the first combination tested by the flight software that exceeds the predefined threshold (generally $4.5\sigma$) is considered the trigger~\citep{GBMBurstCatalog_6Years}. \mod{The full trigger and reporting timeline for GRB~170817A is shown in Table~\ref{table:GBM_timeline}}. GRB~170817A was detected by the GBM flight software on a 256 ms accumulation from 50 to 300 keV ending at 12:41:06.474598 UTC on 2017 Aug 17 (hereafter T0), with a significance of $4.82\sigma$ in the second brightest detector (NaI 2) which, because of the two-detector requirement, sets the threshold.  This value of $4.82\sigma$ does not represent the overall significance of the GRB, but only the significance of the excess for a single detector, for a particular time interval and energy range. Three detectors were above the threshold: NaIs 1, 2 and 5 (cf.~Table~\ref{table:detector_angles}). The significance is calculated as a simple signal-to-noise ratio: excess detector counts above the background model counts divided by expected fluctuations (i.e., the square root of the background model counts). The detection by the flight software occurred 2.4 ms after the end of the data interval. A rapid alert process was initiated, resulting in a GCN\footnote{\url{https://gcn.gsfc.nasa.gov}} Notice being transmitted to observers 14 s later, at 12:41:20 UTC\footnote{\url{https://gcn.gsfc.nasa.gov/other/524666471.fermi}}.
The flight software assigned the trigger a 97\% chance of being due to a GRB, which was reported at 12:41:31 UTC. The initial, automated localizations generated by the GBM flight software and the ground locations had $1\sigma$ statistical uncertainties greater than $20$ degrees but broadly aligned with one of the quadrupole lobes from the skymap produced from the LIGO Hanford (H1) antenna pattern.

\begin{figure}
	\includegraphics[width=15cm]{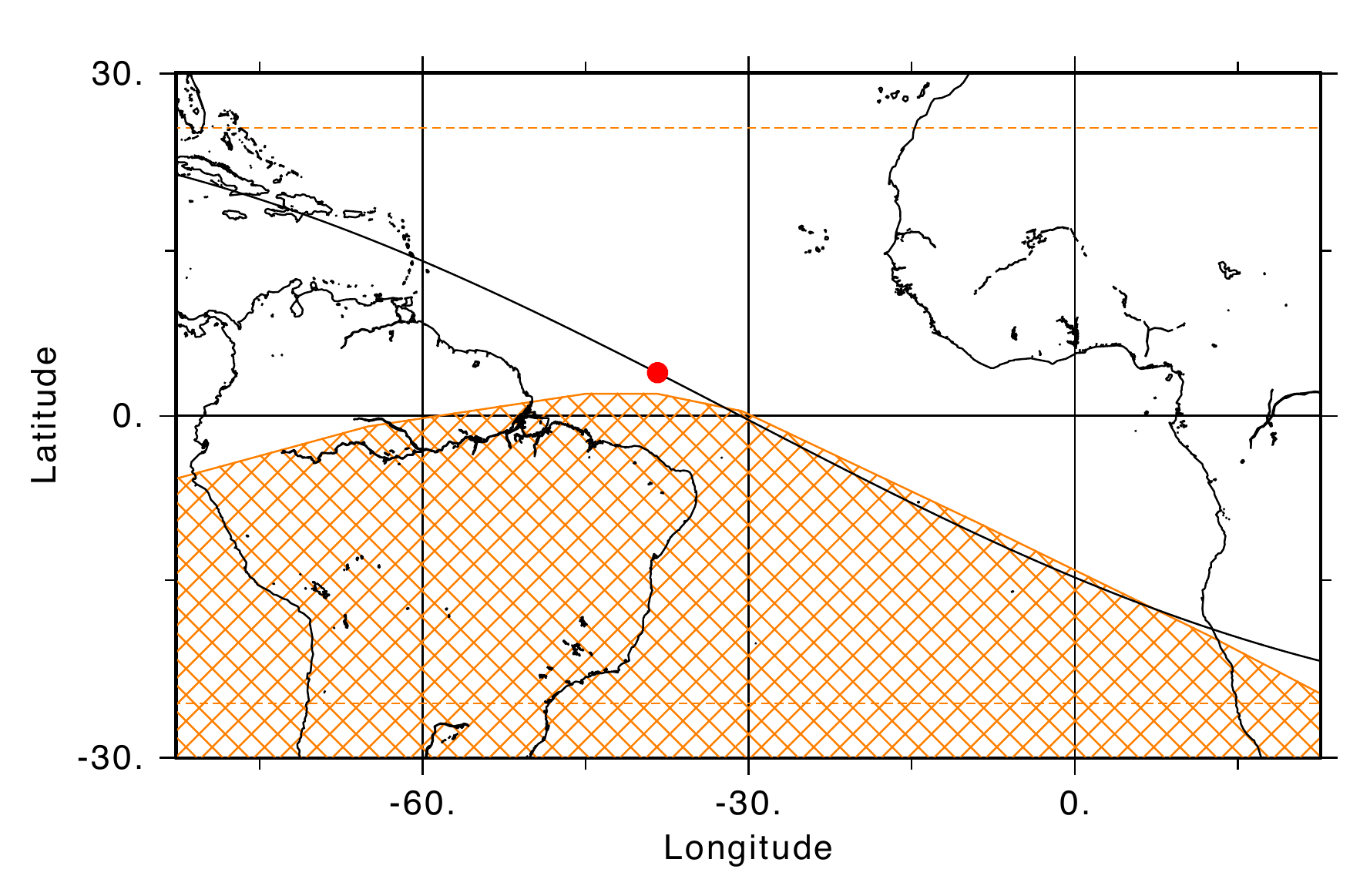}
	\caption{The position of \Fermi at trigger time (red dot) and its orbital path moving from West to East.  The maximum latitudinal extent of \Fermi's orbit is shown by the dashed orange lines and the hatched region is the polygon that defines the  South Atlantic Anomaly region for GBM, inside of which the GBM detectors are turned off.
%
    \label{fig:SAA}}
\end{figure}

The data stream ended 2 minutes post-trigger due to the entrance of \Fermi into the South Atlantic Anomaly (SAA), in which there are high fluxes of charged particles. During passage through the SAA,  high-voltage to the GBM detectors is disabled to extend the lifetime of the detectors, and therefore the detectors cannot collect data. The dependence of the geographical extent of the SAA on the energy of the trapped particles results in different polygon definitions for the GBM and the LAT.  The polygon definition for GBM (see Figure~\ref{fig:SAA}) is slightly smaller than the polygon used for the LAT, enabling the GRB to be detected by GBM while the LAT was turned off and unable to observe it.

\begin{figure}
    \includegraphics[scale=0.50]{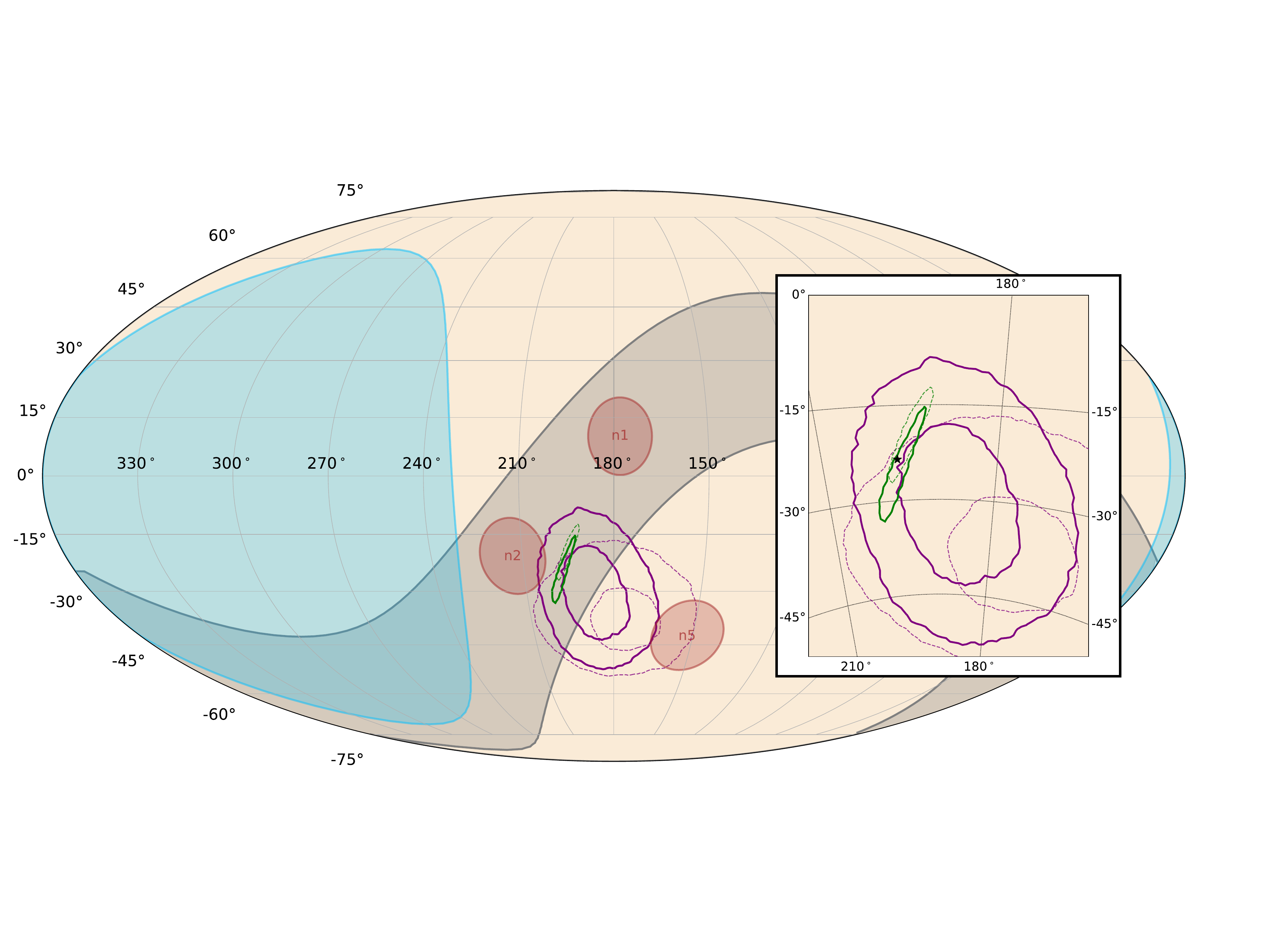}
\caption{The GBM and HLV initial and final localizations. The original GBM human-in-the-loop localization (50\% and 90\% regions) is shown with purple dashed contours, and the original BAYESTAR skymap (90\% region) is shown with green dashed contour. The targeted search localization and the LALInference HLV skymap are the corresponding solid contours. The inset shows a close-up of the GBM localization and the position of the the optical transient candidate (black star). The Earth as seen from \Fermi is shown in blue, the $3\sigma$ IPN annulus is shown as the gray band, and the directions of the three closest NaI detectors are shown in light brown. \label{fig:skymaps}}
\end{figure}

At 13:26:36 UTC, a human-in-the-loop manual localization of GRB~170817A was reported with the highest probability at RA = 180, Dec = $-40$ with a 50\% probability region covering $\sim$500 square degrees and a 90\% probability region of about 1800 square degrees.  This localization in comparison to the HLV localization is shown in Figure~\ref{fig:skymaps}. The large localization region is due to the weak nature of the GRB, the extended tail in the systematic uncertainty for GBM GRB localizations \citep{2015ApJS..216...32C}, and the high backgrounds as \Fermi approached the SAA.  Owing to the importance of this joint detection, an initial circular was sent out at 13:47:37 UTC describing the localization and the event being consistent with a weak short GRB~\citep{gcnGBMGRB170817A_1}. 

\mod{The GBM Team operates a targeted search for short GRBs that are below the triggering threshold of GBM~\citep{Blackburn15}.  This search assumes three different spectral templates for GRBs, each of which are folded through each of the detector responses evaluated over a $1^\circ$ grid on the sky. This method enables a search in deconvolved flux for signals similar to the GRB spectral templates, rather than simpler count-based methods that do not consider the spectrum and detector response.} Guided by detection times from other instruments, such as those by LIGO/{\it Virgo}~\citep{2016PhRvD..93l2003A,PhysRevLett.116.241103,2017PhRvL.118v1101A}, this search requires the downlink of the TTE science data, which can have a latency of up to several hours. Improvements to this search were made in preparation for LIGO's second observing run~\citep{Goldstein16}, which include an un-binned Poisson maximum likelihood background estimation and a spectral template that is more representative of spectrally hard, short GRBs.  The TTE data were transmitted to the ground, and the targeted search completed an automated run at 3.9 hours post-trigger.  The localization from this search is shown in Figure~\ref{fig:skymaps} and is improved relative to the human-in-the-loop localization.  This is primarily due to the improved background estimation provided by the targeted search.  The localization incorporates a $7.6^\circ$ Gaussian systematic uncertainty, determined from processing with the targeted search other short GRBs that triggered GBM on-board and which have accurate locations  determined by other instruments.  The 50\% and 90\% localization credible regions cover $\sim$350 and $\sim$1100 square degrees, respectively.

GBM is also operating an offline untargeted search, which agnostically searches all of the GBM TTE data, for almost all times and directions.  This search runs autonomously and has been publishing candidates since January 2016\footnote{\url{http://gammaray.nsstc.nasa.gov/gbm/science/sgrb_search.html}} and via the GCN since 2017 July 17\footnote{\url{https://gcn.gsfc.nasa.gov/admin/fermi_gbm_subthreshold_announce.txt}}. Similar to the GBM flight software, it searches for statistically significant excesses in two or more NaI detectors.  Compared to the flight software, it has a better background model and tests more time intervals.  The untargeted search, in its standard form, did not detect GRB~170817A.  This is because the untargeted search has several tests to ensure quality background fits in order to avoid spurious candidates.  These tests cause $\approx 2$\% of the TTE data to be omitted from the search.  One of the tests rejects data intervals with rapidly changing background rates, as sometimes occurs near the SAA as \Fermi moves into/away from high trapped particle fluxes.  This test terminated the search at 12:40:50, 16 s before the GRB.  Relaxing this standard test, a good background fit is obtained by the program and the GRB is found at high significance and is classified as highly reliable because more than two detectors had significant excesses. In the 320 ms detection interval, NaIs 1, 2, 4, 5 and 11 were found to have significances of $5.63\sigma, 5.67\sigma, 3.41\sigma, 6.34\sigma$ and $3.57\sigma$, respectively.  The signal in NaI 11 is due to viewing GRB photons scattered from the Earth's atmosphere and viewing the GRB through the LAT radiator and the back of the detector (which gives a larger response then viewing perpendicular to a source).

\section{Standard GBM Analysis}
\label{sec:StandardAnalysis}
As part of GBM operations, all triggered GRBs are analyzed following standardized procedures, and results from these analyses are released publicly in the form of catalog publications~\citep{GBMSpectralCatalog_4Years,GBMBurstCatalog_6Years} and a searchable online catalog hosted by the HEASARC\footnote{\url{https://heasarc.gsfc.nasa.gov/W3Browse/fermi/fermigbrst.html}}.  Additionally, all triggered data files are publicly available soon after the data are downlinked from the spacecraft and processed automatically in the ground pipeline.  In this section, we present the result of the standardized analysis for GRB~170817A so that it may be placed in context of other GRBs that trigger GBM.

The response of the GBM NaI detectors is strongly dependent on the angle between the detector and source location, with additional contributions from scattering from the spacecraft and off the Earth's atmosphere. Due to this, a source position must be assumed for the GRB so that detector responses can be generated, mapping incident photon energies to observed count energies.  In all following analysis, we assume the position of the optical transient candidate. For standard analysis, we use the NaI detectors that have observing angles to the source position $\leq 60^\circ$ since the response is reduced beyond this angle, and the tradeoff of the low response with possible systematics is poorly understood.  Although the BGO response does not depend as strongly on viewing angle as do the NaI detectors, the BGO detector with the smallest viewing angle to the source is used. Additionally, detectors are not used for analysis if portions of the spacecraft or LAT block the detector from viewing the source.  The detector angles and detectors selected for analysis are shown in Table \ref{table:detector_angles}.

\subsection{Duration}
The duration of GRBs is usually defined by the \Tninety, which is the time between reaching 5\% and 95\% of the cumulative observed fluence for the burst in the canonical observing energy range of 50-300 keV. Because the orientation of \Fermi may change with respect to the source position over the duration of a GRB,  the GBM \Tninety calculation is performed on a photon spectrum rather than the observed count spectrum. This removes the possibility of bias owing to the changing response of the detector to a changing source angle, an effect that is most important for long GRBs. This is the standard method for all GBM \Tninety calculations although other techniques exist.  A power-law spectrum with an exponential cut-off is fit to the background-subtracted data over a time interval that begins prior to the trigger time of the burst and extends well beyond its observed duration, using the detector response for the best available source position. The fits are performed sequentially over 64 ms (1.024 s) time bins for short (long) GRBs.  Either side of the impulsive GRB emission, the presence of stable and long-lived plateaus in the deconvolved time history indicates the times at which 0\% and 100\% of the burst fluence has been recorded, and the 5\% and 95\% fluence levels and their associated times are measured relative to these plateaus to yield the \Tninety duration. In addition to the \Tninety, this analysis produces an estimate of the peak flux and fluence in the standard GBM reporting range of 10--1000 keV. 

Following the recipe of the standard analysis, we use detectors NaI 1, 2, and 5 to estimate the \Tninety.  A polynomial background is fit to 128-channel TTE data, binned to 8 energy channels in each detector.  We find the \Tninety to be  $2.0 \pm 0.5$ s, starting at  T0$-0.192$ s.  We note that there appears to be emission below the 50--300 keV energy range after $\sim0.5$ s which contributes to the deconvolution of the spectrum during that time, thereby extending the \Tninety beyond what is strictly observed in 50--300 keV (cf.~Figure~\ref{fig:BTTE_Chan_NaI}). For GRB~170817A, the peak photon flux measured on the 64 ms timescale and starting at T0 is $3.7 \pm 0.9 \ \rm ph \ s^{-1} \ cm^{-2}$.  The fluence over the \Tninety interval is $(2.8 \pm 0.2)\times 10^{-7} \ \rm erg \ cm^{-2}$.  

\subsection{Spectrum}
A standard spectral analysis is performed for each triggered GRB and the results are included in the GBM spectral catalog.  Two lightcurve selections are performed: a selection over the duration of the burst, and a selection performed at the brightest part of the burst.  The first selection is performed by combining the lightcurves of the NaI detectors, identifying regions that have a SNR $\geq 3.5$, and applying those signal selections to each detector individually.  This permits a time-integrated spectral fit of regions that are highly likely to be true signal with minimal background contamination.  The second lightcurve selection is performed by summing up the same NaI detectors and selecting the single brightest bin---for short (long) GRBS the brightest 64 ms (1024 ms) bin---and the selection is applied to all detectors individually. For both the time-integrated and peak spectra, the data from each detector are jointly fit via the forward-folding technique using RMfit\footnote{\url{https://fermi.gsfc.nasa.gov/ssc/data/analysis/rmfit}}.  \mod{Specifically, minimization is sought for the Castor C-statistic~\citep{cstat} using the Levenberg-Marquardt non-linear least-squares minimization algorithm.}  Further details on standard spectral fitting analysis procedures and selections are given in~\citet{goldstein12}. The fit results for GRB~170817A are shown in Table~\ref{tab:Spectra}.

The time-integrated selection produces a 256 ms time interval from T0$-0.192$ s to T0$+0.064$ s and is statistically best fit by an exponentially cut-off power law, \mod{which is referred to as a Comptonized spectrum in the GBM spectroscopy catalog (see Eq. 3 in~\citet{GBMSpectralCatalog_4Years}).}  This fit results in a weakly-constrained power-law index of $0.14 \pm 0.59$ and a break energy, \mod{characterized as the $\nu F_\nu$ peak energy}, \Epeak$=215 \pm 54$ keV.  The averaged energy flux over this interval in 10--1000 keV is $(5.5 \pm 1.2)\times 10^{-7} \ \rm erg \ s^{-1} \ cm^{-2}$, and the corresponding fluence is $(1.4 \pm 0.3)\times 10^{-7} \ \rm erg \ cm^{-2}$.  Note that the fluence over the \Tninety interval is larger compared to this interval due to the fact that this time interval is considerably shorter than the \Tninety.

The 64 ms peak selection from T0$-0.128$ s to T0$-0.064$ s is also statistically best fit by a Comptonized spectrum.  Again the parameters are poorly constrained with the power-law index $=0.85 \pm 1.38$ and \Epeak$=229 \pm 78$ keV.  The resulting peak energy flux from this spectral fit in 10--1000 keV is $(7.3 \pm 2.5)\times 10^{-7} \ \rm erg \ s^{-1} \ cm^{-2}$.

\subsection{Comparison to the GBM Catalogs}
\begin{figure}
    \includegraphics[scale=0.7]{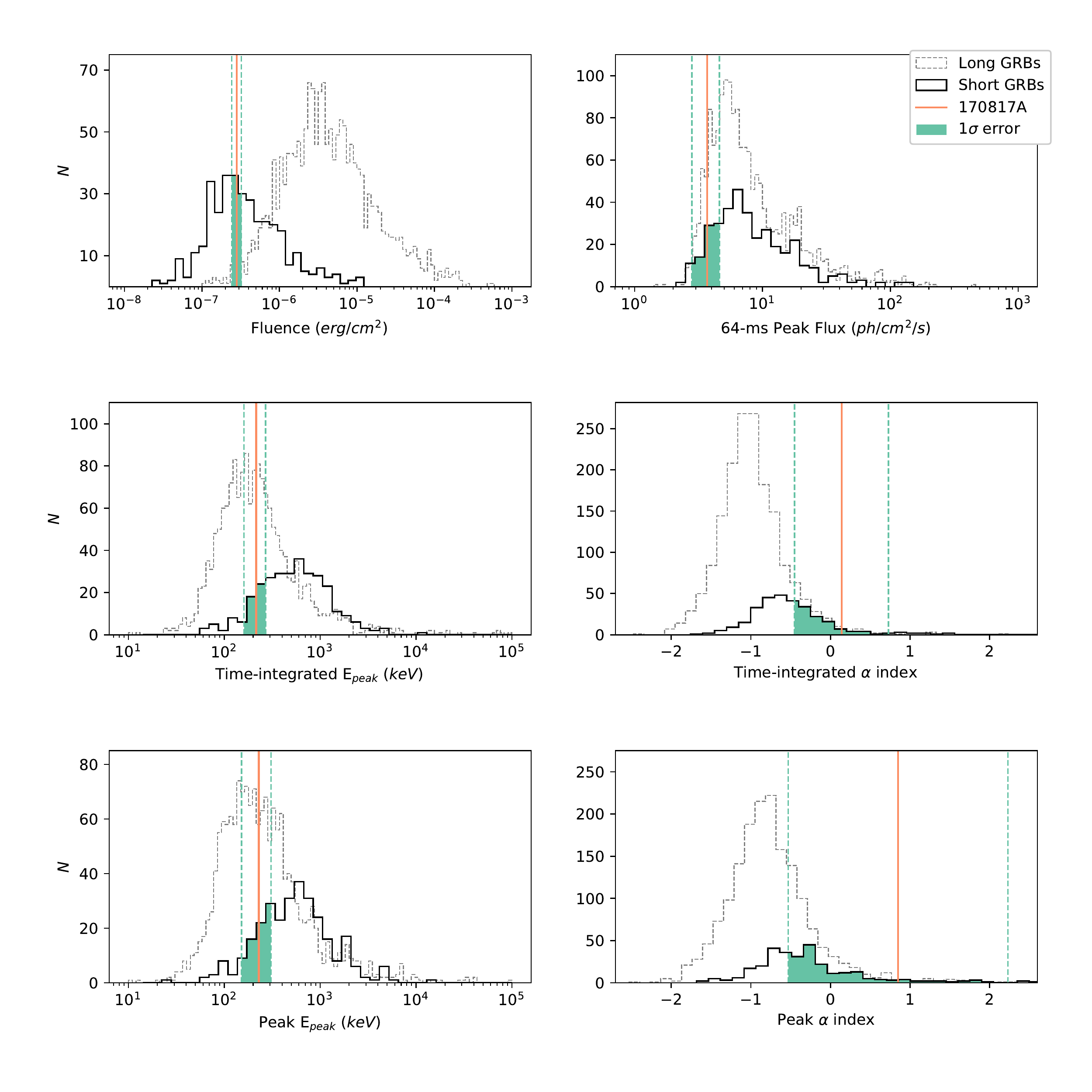}
	\caption{Distributions comparing measured and spectral fit parameters of GRB~170817A to those of short (black) and long (grey) GBM GRBs. The value obtained for GRB~170817A is indicated by the orange line, and the $1\sigma$ uncertainty is both shaded and delineated in green. Both the fluences (top left) and 64 ms peak fluxes (top right) were calculated in the 10--1000 keV range. The time-integrated spectral parameters (middle left and right) and the peak spectral parameters (bottom left and right) were computed using the Comptonized spectrum.}
	\label{fig:spectral_comp}
\end{figure}

We compare the standard analysis of GRB~170817A to other GRBs contained in the GBM Burst Catalog~\citep{GBMBurstCatalog_6Years} and Spectral Catalog~\citep{GBMSpectralCatalog_4Years}. A GBM catalog of time-resolved spectroscopy has also been produced~\citep{GBMTimeResolvedSpectralCatalog}, but this GRB is too weak to perform the required time-resolved spectral fits to compare to that catalog.
Keeping to the traditional definition of short and long GRBs, our sample comprises of 355 short bursts and 1714 long bursts spanning the beginning of the mission to 2017 August 27. For GRB~170817A, we compare the 64 ms peak photon flux and the fluence obtained from the duration analysis. The distributions are shown in Figure \ref{fig:spectral_comp}, as are the distributions of the cut-off power law parameters for the time-integrated and peak spectra. 

Because short GRBs are typically defined as those with duration $< 2$ s, they generally have lower fluences than those of long bursts. The fluence for GRB~170817A is consistent with those obtained for short GRBs, falling within the 40th--50th percentile of the short distribution. For the 64 ms peak photon flux, the long and short distributions are similar, with median values of 6.56 and 7.26  $\rm ph \ s^{-1} \ cm^{-2}$, respectively.  GRB~170817A, in comparison, lies at the $\sim 10$th percentile of both distributions and is thus weaker than the average GRB on that timescale. \mod{As observed in GBM,} short bursts tend to have higher \Epeak values than long bursts.  For both time selections, the \Epeak of GRB~170817A falls at the $\sim 15$th percentile of the short GRB distribution, corresponding to the softer tail, and near the median of the long GRB distribution. Long GRBs display a median lower power-law index of $-1.01$, while the short GRBs have a slightly harder index with a median of $-0.58$ for the time-integrated distribution and $-0.27$ for the peak distribution. The power-law index for GRB~170817A, though weakly constrained, lies within the positive tail of the distributions between the 85th--95th percentiles for long and short GRBs.

\section{Classification}
\begin{figure}
	\subfigure[]{\label{fig:t90_distrib}\includegraphics[scale=0.45]{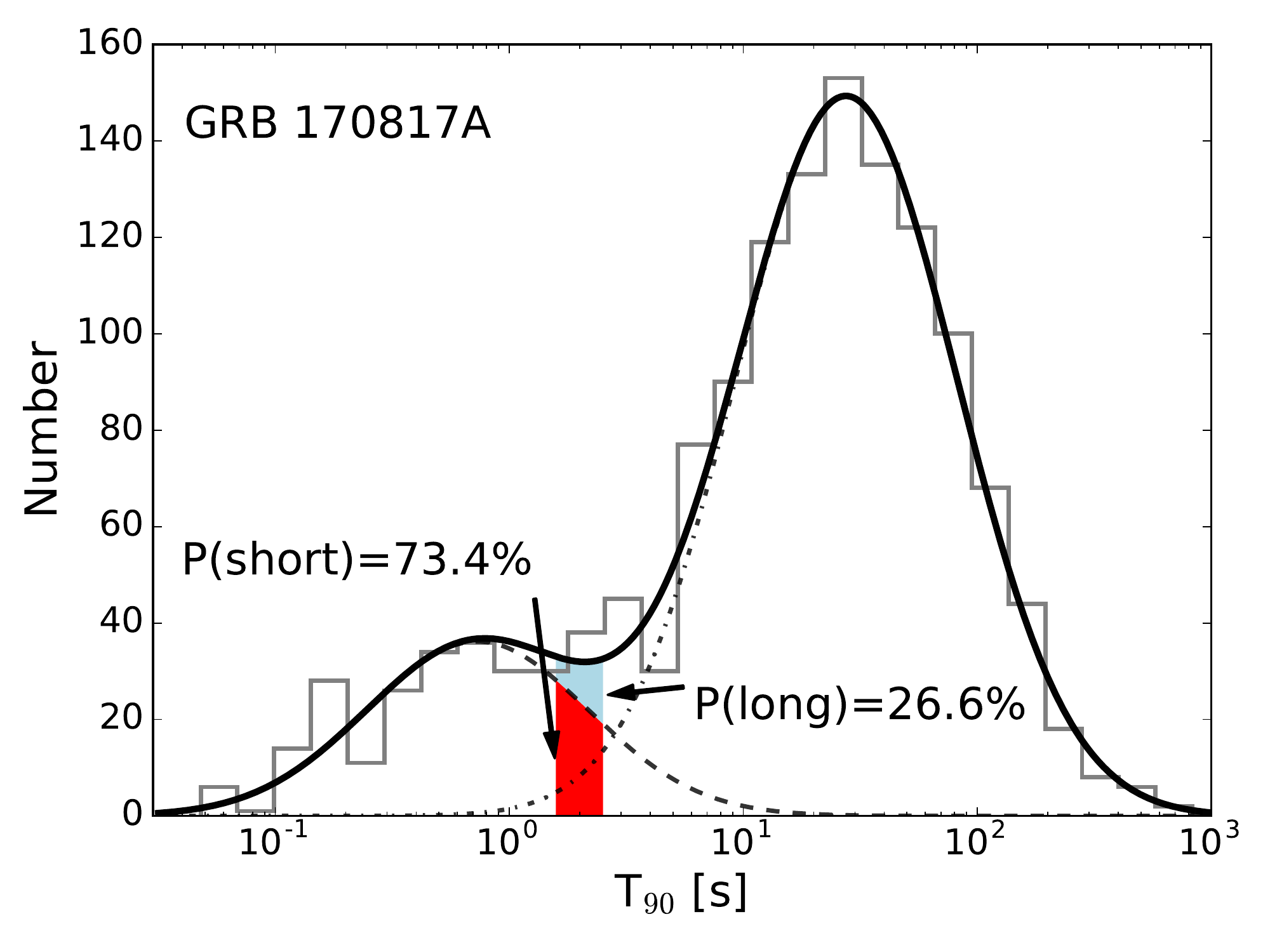}}
	\subfigure[]{\label{fig:hardness_t90}\includegraphics[scale=0.37]{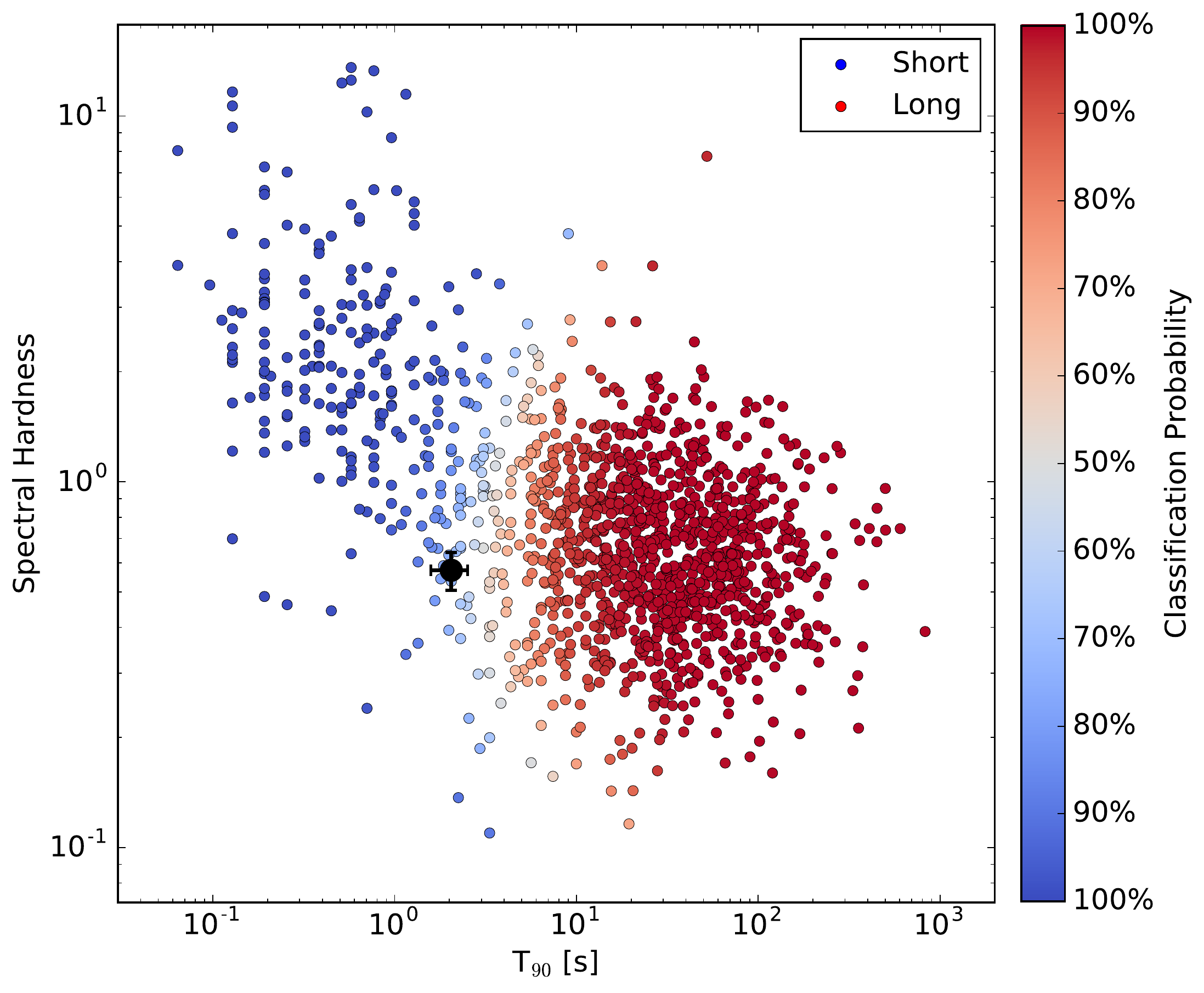}}
	\caption{{\it [Left]} The GBM \Tninety distribution fit with two log normal distributions. The $1\sigma$ confidence interval for GRB~170817A is shaded below the summed curve. The red region is the probability that the event belongs to the short class, while the light blue is the probability that it belongs to the long class. {\it [Right]} The duration (\Tninety) vs the hardness ratio, an analog for the spectral hardness of the burst. Assuming exactly two distinct populations the data are fit with two-dimensional log-normal distributions. Red dots are those most likely to belong to the long class, and blue dots to the short class. The black cross is the centroid and $1\sigma$ uncertainty for GRB~170817A.\label{fig:Classification}}
\end{figure}

Historically, GRBs have been classified based on their duration. During the BATSE era, the duration distribution in the 50--300 keV band, plotted as a histogram, showed evidence for bimodality; the shorter population, peaking at $\sim 1$s duration was termed `short' while the longer and more dominant population peaking at $\sim 30$ s duration was termed `long'. The overlap of the two original distributions in BATSE at $\sim 2$ s was designated as the classification boundary between the two GRB types~\citep{1993ApJ...413L.101K}, although the accumulation of more GRBs has shown that the overlap of the two distributions has changed and can be affected by the energy band over which the duration is estimated. Figure~\ref{fig:t90_distrib} shows the \Tninety duration distribution of GRBs that triggered GBM through 2014 July 11 \citep{GBMBurstCatalog_6Years}. The \Tninety for GRB~170817A is shown relative to the distribution of the GBM GRBs, and when the distributions are modeled as two log-normals, the probability that the GRB belongs to the short class is $\sim73$\%. 

Short-duration GRBs were also observed to be spectrally harder than the average long GRB~\citep{FirstHardnessRatioPaper}.  One way to represent this distinction is to calculate a hardness ratio, which is the ratio of the observed counts in 50--300 keV compared to the counts in the 10-50 keV band and is useful in estimating the spectral hardness of an event without the need to perform deconvolution and fitting a spectrum. Figure~\ref{fig:hardness_t90} shows the hardness--duration plot revealing the two distinct populations of short--hard GRBs and long--soft GRBs.  Similar to the modeling of the \Tninety distribution, the hardness-duration can be modeled as a mixture of two-dimensional log-normal distributions.  The location of GRB~170817A is shown on this diagram, and using the mixture model, we estimate the probability that it belongs in the short--hard class as $\sim$72\%.

Two types of progenitors have been proposed for these two GRB classes: collapsars as the progenitors for long GRBs~\citep{MacFadyenCollapsar}, and the compact binary mergers as the progenitors for short GRBs~\citep{eichler1989nucleosynthesis,fox2005afterglow,d2015short}.  The connection between long GRBs and collapsars is well-established; however the connection of short GRBs and mergers has been only circumstantial.  Owing to the fast transient nature of the prompt emission, the rapid fading of the afterglow emission, and the typical offset from the putative host galaxy, a firm connection between a short GRB and its theoretical progenitor required a coincident GW signal.

\section{Detailed Analysis}
\label{sec:DetailedAnalysis}
In addition to the standard analysis that is performed on each GBM-triggered GRB, we include a more detailed analysis of this GRB: investigating the spectrum in different time intervals, estimating the spectral lag properties of this burst, estimating the minimum variability time, and commenting on possible periodic or extended emission.  Where applicable, the following analyses employ an improved background estimation technique for weak signals--the same background method used in the targeted search~\citep{Goldstein16}.  This background estimation provides a standardized method that does not rely on user selections of background regions, and models the background in each energy channel independently without assuming an approximating polynomial shape of the background.

\subsection{Spectral Analysis}\label{sec:spectra}
\begin{figure}
	\includegraphics[width=15cm]{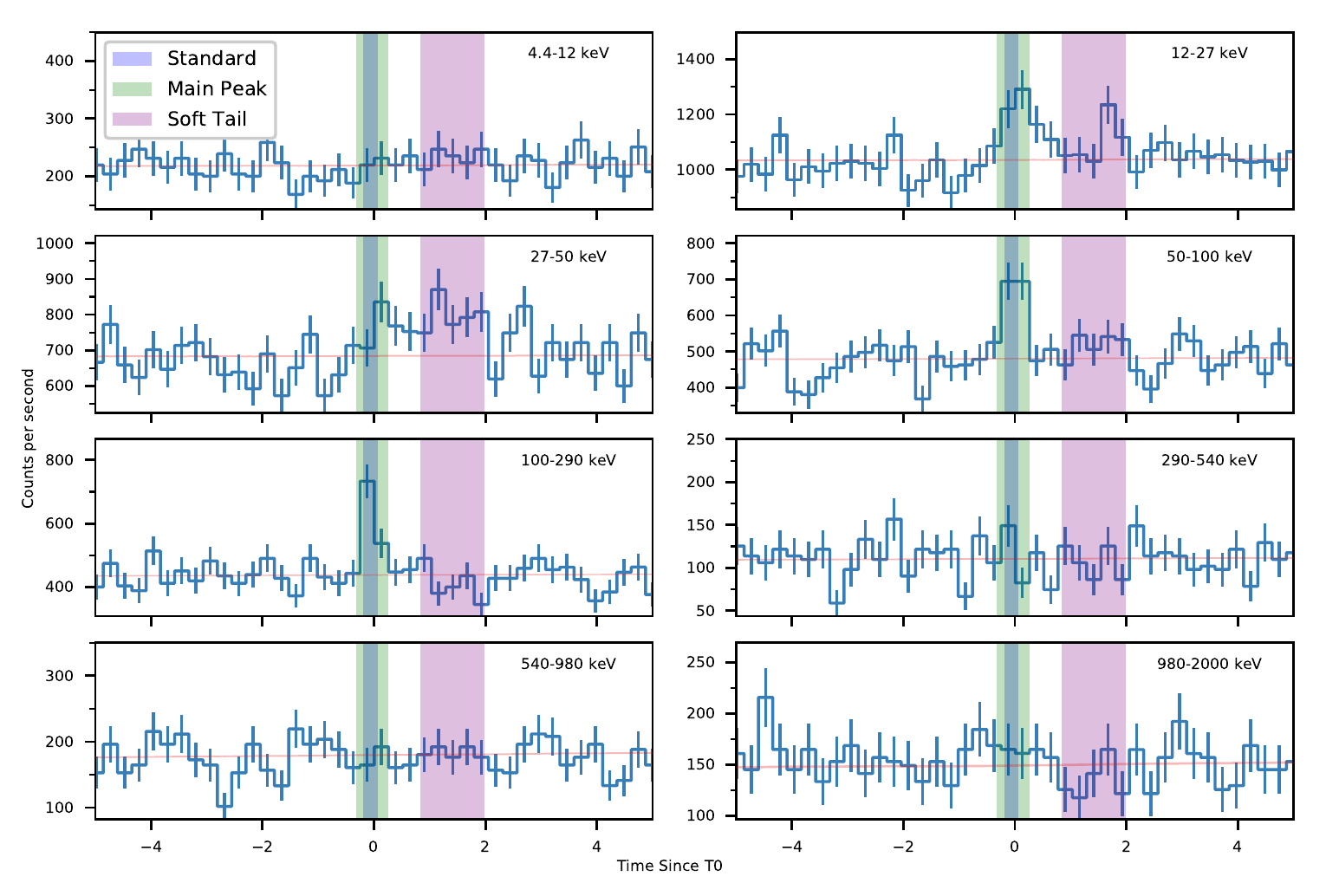}
	\caption{The 256 ms binned lightcurve of GRB~170817A for NaI 1, 2, and 5 over the standard 8 CTIME energy channels. The shaded regions are the different time intervals selected for spectral analysis.  The inclusion of the lower energies shows the soft tail out to T0+2 s.}
    \label{fig:BTTE_Chan_NaI}
\end{figure}

\begin{figure}
	\includegraphics[width=15cm]{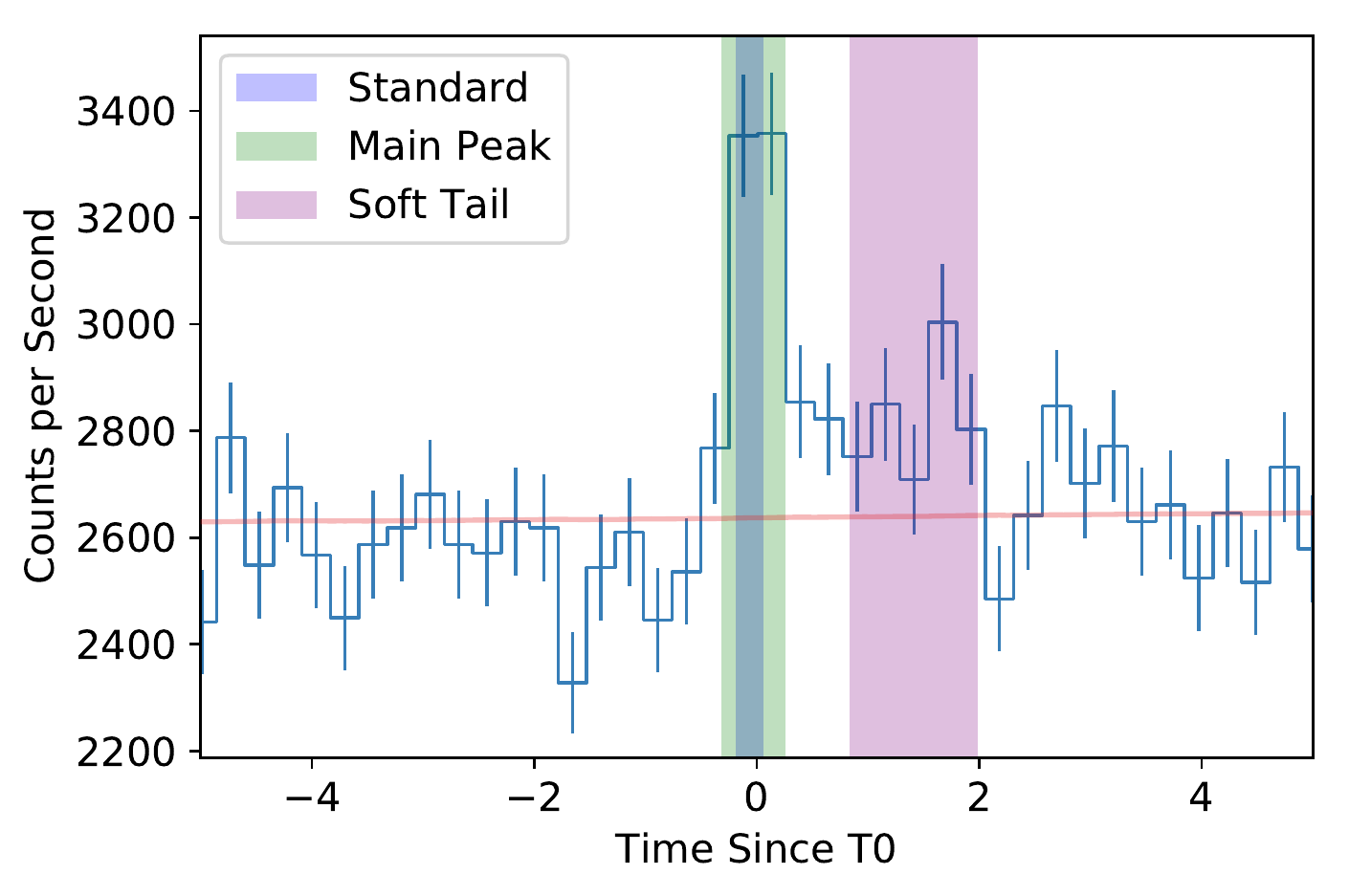}
	\caption{The 256 ms binned lightcurve of GRB~170817A in the 10--300 keV band for NaI 1, 2, and 5. The shaded regions are the different time intervals selected for spectral analysis.  The inclusion of the lower energies shows the soft tail out to T0+2 s.}
	\label{fig:BTTE_sum_10to300}
\end{figure}

\begin{figure}
	\subfigure[]{\label{fig:mainpulsefit}\includegraphics[scale=0.45]{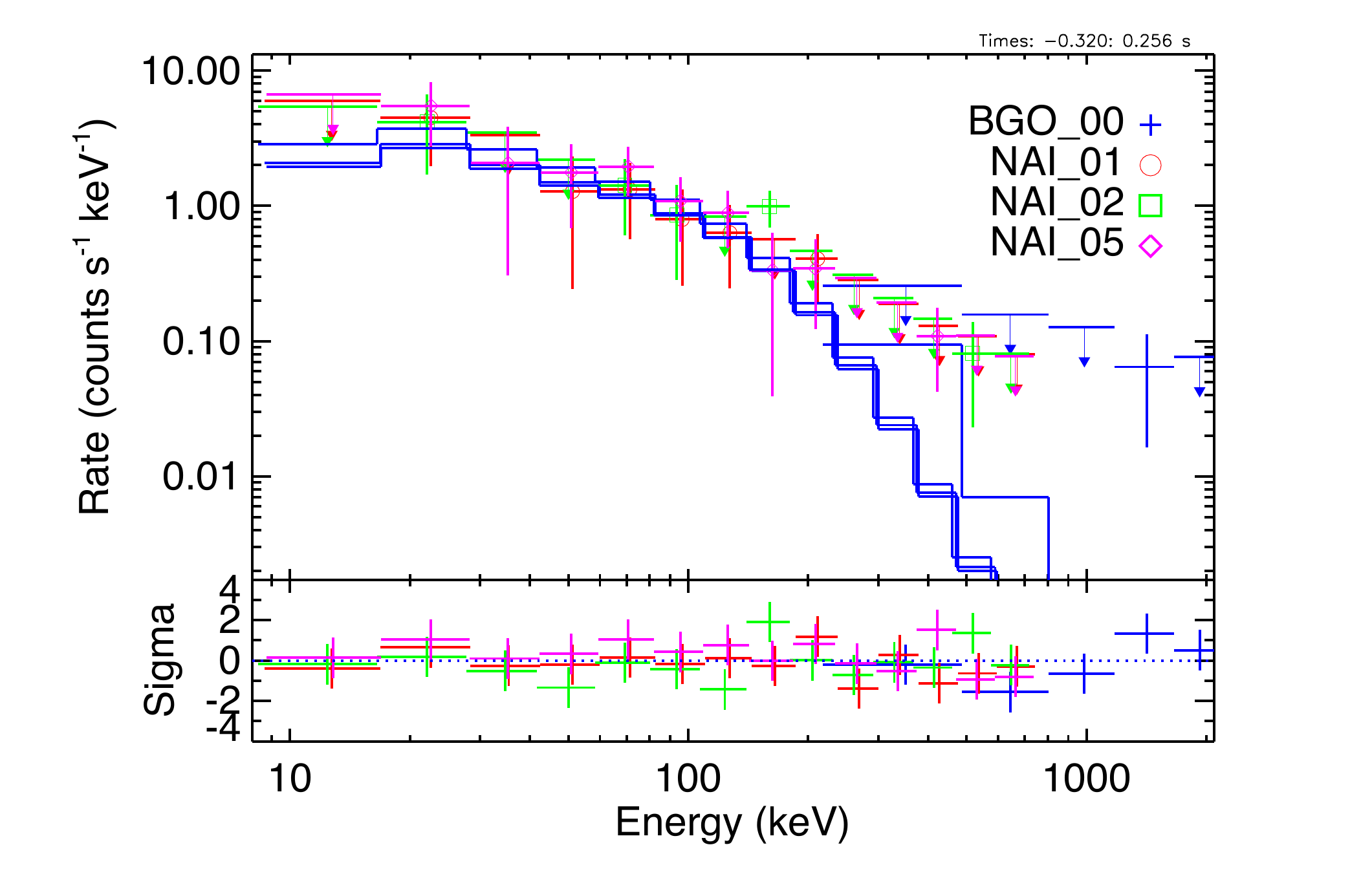}}
	\subfigure[]{\label{fig:softfit}\includegraphics[scale=0.45]{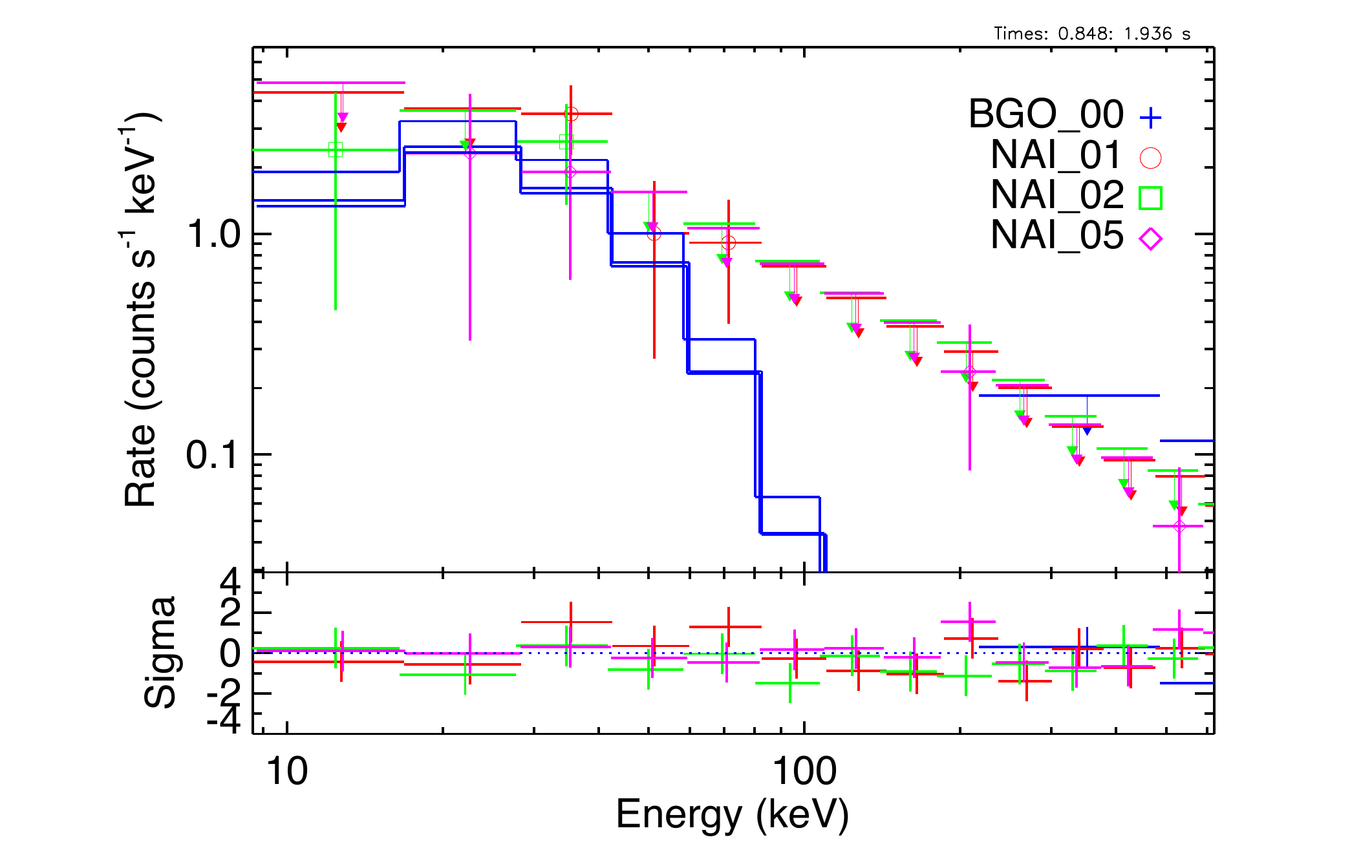}}
	\caption{Spectral fits of the count rate spectrum for the {\it [Left]} main pulse (Comptonized) and {\it [Right]} softer emission (black body).  The blue bins are the forward-folded model fit to the count rate spectrum, the data points are colored based on the detector, and $2\sigma$ upper limits estimated from the model variance are shown as downward-pointing arrows.  The residuals are shown in the lower subpanels.   
    \label{fig:Spectral_Fits}}
\end{figure}

After visual inspection of the lightcurve (shown in Figures~\ref{fig:BTTE_Chan_NaI} and~\ref{fig:BTTE_sum_10to300}), we first select the main pulse from T0-0.320 s to T0+0.256 s for spectral analysis.  We perform the spectral analysis in RMfit with a background model created from the un-binned Poisson maximum likelihood background estimation. This interval is best fit by a Comptonized function with \Epeak$=185 \pm 62$ keV, $\alpha=-0.62\pm 0.40$, and the resulting time-averaged flux is $(3.1 \pm 0.7)\times 10^{-7} \ \rm erg \ s^{-1} \ cm^{-2}$. The fit to the count rate spectrum is shown in Figure~\ref{fig:mainpulsefit}. We compare this model to the best-fit power-law over the same interval, resulting in a power-law index of $-1.48$. By performing 20,000 simulations assuming the power law as the true source spectrum, we find that the C-stat improvement of 10.6 as observed for the cut-off power law corresponds to a chance occurrence of $1.1\times 10^{-3}$.  \mod{Therefore, we conclude that the Comptonized function is statistically preferred over the simple power law.}

As can be seen in Figures~\ref{fig:BTTE_Chan_NaI} and~\ref{fig:BTTE_sum_10to300}, the main pulse of the GRB appears to be followed by a weak and soft emission episode. It is not immediately clear if it belongs to the GRB or if it is due to background variability. To ascertain the connection of the soft emission to the main pulse, we localize this soft excess using the standard GBM localization procedures, using the 10--50 keV data and a soft spectral template devised for the localization with good statistics of non-GRB transients with softer spectra, such as magnetars or solar flares.  We find that the soft emission localizes to RA=181, Dec=-30 with the 50\% (90\%) credible region approximately circular with a radius of 15 (28) degrees, in good agreement with both the localization of the main pulse and the HLV sky map.

In addition to localizing the softer emission, a Bayesian block method was used to analytically determine whether the longer softer emission could be significantly detected. The algorithm~\citep{Scargle2013}  characterizes the variability in the TTE data by determining change points in the rate, thereby defining time intervals (called ``blocks'') of differing rates.  This method can be used to test for separate statistically significant signals against the Poisson background. The algorithm has previously been used extensively to evaluate Terrestrial Gamma-ray Flash (TGF) candidates found in off-line searches of the GBM TTE data~\citep{Fitzpatrick2014,Roberts2017}. 

Initially, the TTE data for NaI detectors 1,2 and 5 were investigated $\pm$5 seconds on either side of the GBM trigger time of GRB~170817A and analyzed using a false positive probability (p$_{0}$) of 0.05, previously determined to be a good value from studies by~\citet{Scargle2013}. We find the Bayesian block duration using all three detectors to be 0.647~s, however when running the analysis again using just NaI 2 (the detector with the best source-detector geometry), some softer emission after the initial pulse is deemed significant enough by the algorithm to extend the duration time to 1.12~s. This soft emission after the main pulse is not deemed significant when the algorithm is used separately over the data from NaI 1 and 5. One possible explanation for this is that the effective area of NaI 1 and 5 are $\sim 20-25$\% lower compared to NaI 2 for soft emission (see~\citet{2009Meegan,Bissaldi2009} for effective area dependence on source--detector angle).


We find that the spectrum of the soft emission from T0+0.832 to T0+1.984 is well fit by a blackbody with a temperature $kT=10.3\pm1.5$ keV (see Figure~\ref{fig:softfit} for the spectral fit). The blackbody fit has an improvement in C-stat of 18 compared to a power law fit (same number of degrees of freedom), and we find that it is statistically significant, at the $< 1\times10^{-4}$ level, via simulations. Assuming that the blackbody is the true spectrum of the soft emission, the fluence of the soft emission is $\sim$34\% of the main pulse (10--1000 keV range). The results of these spectral fits are listed in Table~\ref{tab:Spectra}.  We also attempted to fit a Comptonized function, which approximates the shape of the blackbody with \Epeak$=38.4 \pm 4.2$ keV and an unconstrained power-law index of $4.3\pm3.0$.  \mod{The large uncertainty in the power-law index is likely due to the fact that the \Epeak is near the low-energy end of the NaI observing band, so there are not enough energy channels to constrain the power-law index.}  The improvement in C-stat of 2 units for the additional degree-of-freedom does not indicate that the Comptonized is statistically preferred.

\mod{If we assume that this softer emission is indeed thermal, this pulse may be explained as photospheric emission from a cocoon. Postulated cocoon emission is ubiquitous in the case of collapsars~\citep{Peer+06cocoon,Nakar+17cocoon}, and may also be present in the binary neutron star merger scenario. In this picture, significant energy is deposited by the jet responsible for the GRB in the surrounding dense material (e.g. in the debris disk)~\citep{ramirezruiz02}. This results in a cocoon that expands until it achieves a mildly relativistic, coasting Lorentz factor, $\Gamma_c\sim $few, according to the same dynamics as GRB jets \citep{Meszaros+93gasdyn}. Cocoon emission, however, subtends a wider opening angle making it essentially isotropic. A $kT\approx10$ keV temperature blackbody spectrum is in agreement with expectations from such a scenario \citep{Lazzati+17cocoon}.  Furthermore, the $T_{\rm soft}\approx1$ s duration of the soft pulse can be related to the typical angular timescale (assuming it is longer than the diffusion timescale at the start of the cocoon expansion), yielding an emission radius $R_{\rm phot,c}\approx 10^{12} ~{\rm cm}~(\Gamma_{\rm c}/4)^2 (T_{\rm soft}/1 ~{\rm s})$ that is also broadly consistent with expectations from a cocoon scenario.}

\subsection{Spectral Lag}
Spectral lag, the shift of the low-energy lightcurve for a GRB compared to a higher-energy lightcurve is a well-known observed phenomenon exhibited in GRBs~\citep{FenimoreLags}.  Long GRBs typically have a soft lag, where the low-energy lightcurve lags behind the high-energy lightcurve.  Short GRBs, due to their shorter timescale and generally lower fluence, have spectral lags that are more difficult to measure.  Many short GRBs are consistent with zero lag~\citep{BernardiniLags}, while some are consistent with soft lag, and others are consistent with hard lag (high-energy lightcurve lags low energies)~\citep{YiLags}.  \mod{There are a number of proposed explanations for the observed spectral lag. Among the likely explanations are effects from synchrotron cooling~\citep{KazanazLags} and kinematic effects due to observing the GRB jet at a large viewing angle~\citep{SariPiran,SalmonsonLags,ChuckCurvature}, both of which can manifest as observed spectral evolution of the prompt emission in the jet~\citep{BandLags,KocevskiLiangLags}.}

Several methods have been devised to estimate the spectral lag.  We choose to use the discrete cross-correlation function (DCCF) as defined in~\citet{BandLags} to measure the correlation between the lightcurve in two different energy bands. The DCCF has values that typically range from $-1$ (perfect anti-correlation) to +1 (perfect positive correlation).  The general method is to shift one lightcurve relative to the other lightcurve, with each time shift discretized as a factor of the binning resolution.  At each time shift, the DCCF is computed.  \mod{The DCCF as a function of the time shift should peak when the correlation between the two lightcurves reaches the maximum. If this maximum is $< 1$, it could be due to different effects, particularly the brightness of the lightcurve relative to the background or intrinsic physics of the source that causes significant differences in the lightcurve at different energies.} In order to utilize multiple detectors, we account for the estimated background in each detector, combining the background uncertainties into the calculation of the DCCF.  Sometimes a second-order polynomial is fit to the DCCF to find the maximum; however this is inadequate when a lightcurve contains many pulses, when the signal is relatively weak, or if the identification of the signal is not precise.  Therefore, to find the maximum of the DCCF, \mod{we estimate the trend of the DCCF using non-parametric regression~\citep{lowess}. Because the regression produces no functional form, we perform quadratic interpolation of the regression between the evaluated data points and determine where the regression is at maximum.} To estimate the uncertainty, we create Monte Carlo deviates of the DCCF and fit using the same method. The median and credible interval can then be quoted for the spectral lag.

\begin{figure}
	\subfigure[]{\label{fig:lagEnergy}\includegraphics[scale=0.45]{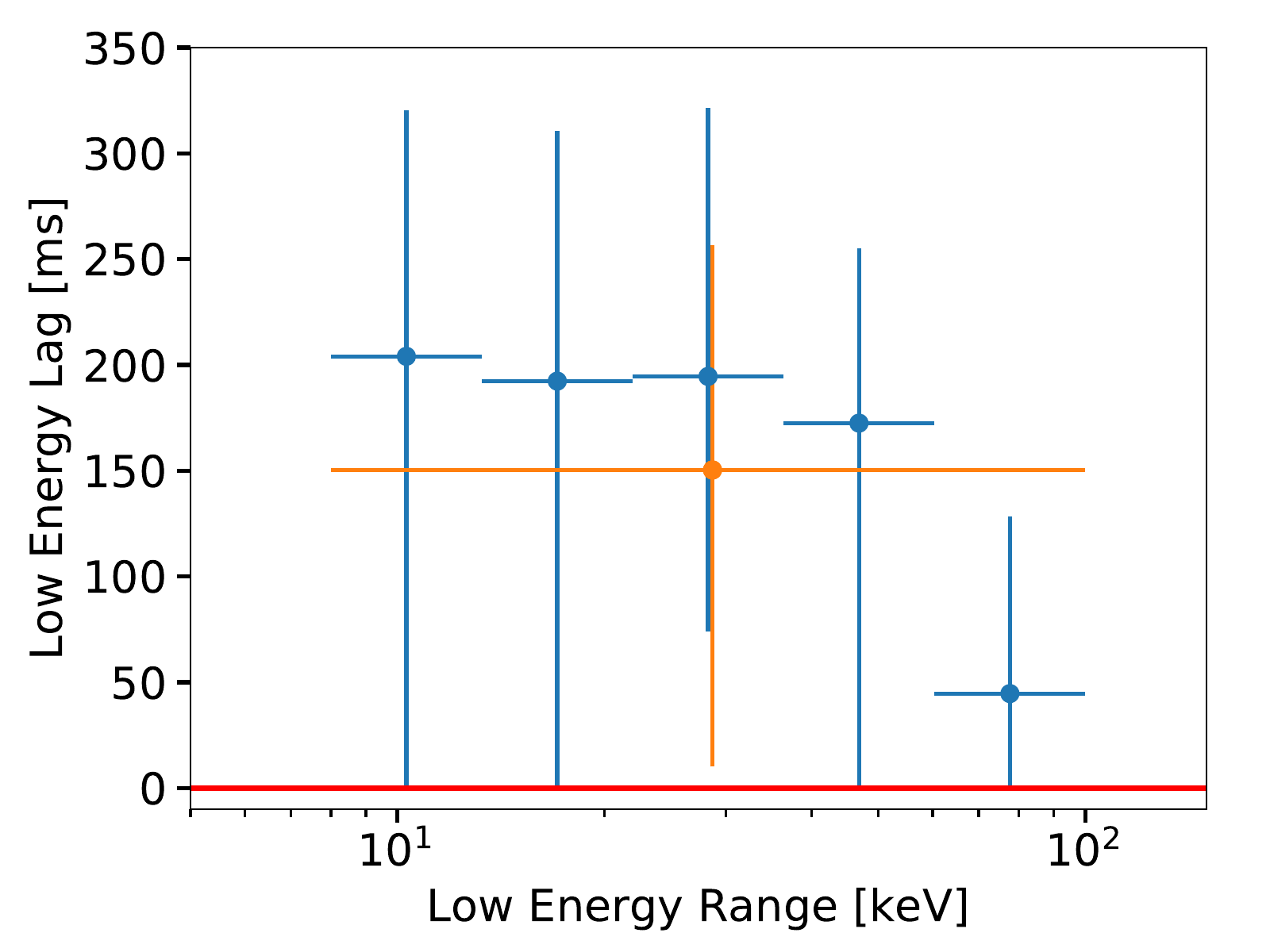}}
	\subfigure[]{\label{fig:lagCCF}\includegraphics[scale=0.45]{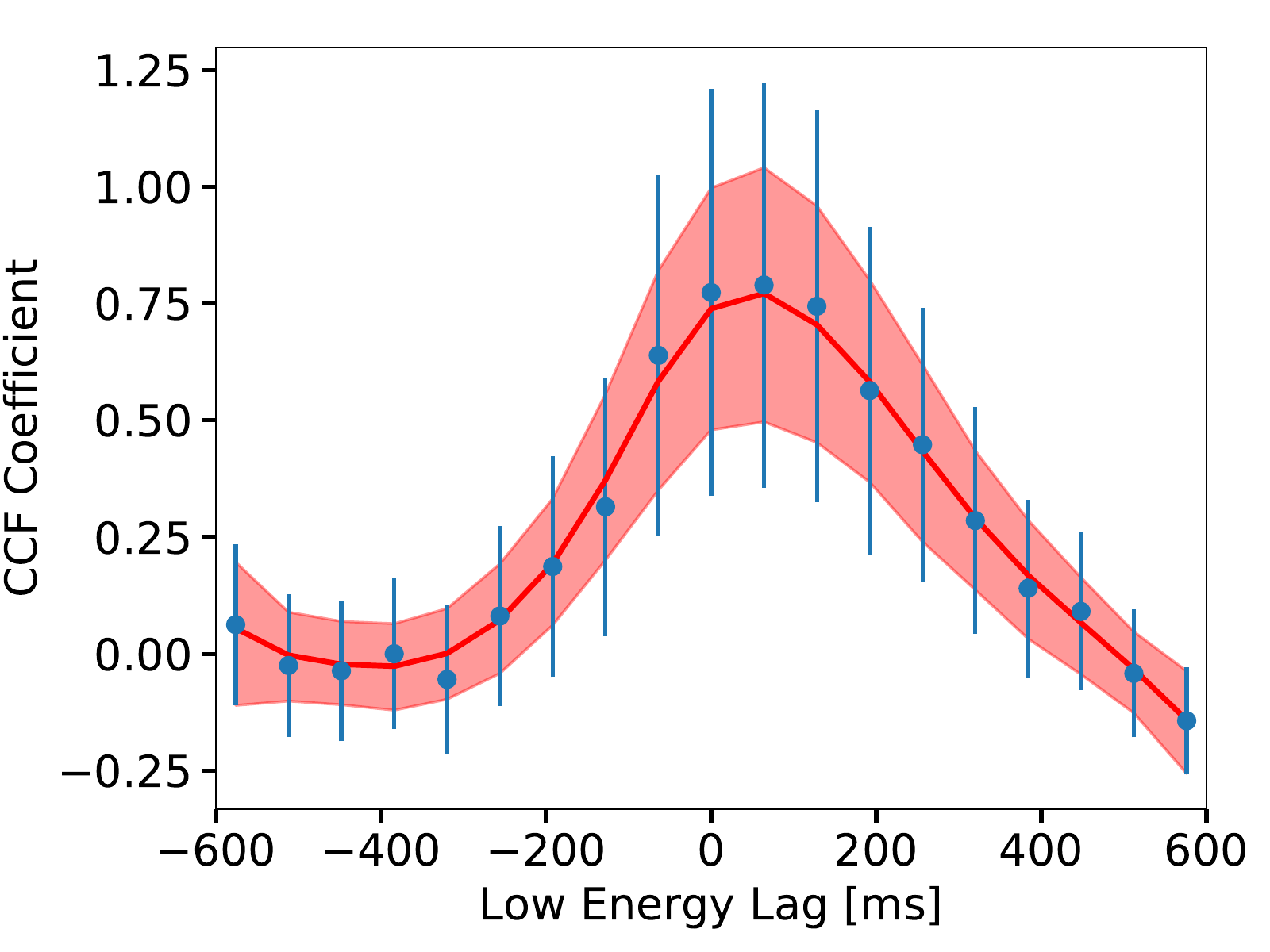}}
	\caption{{\it [Left]} The estimated lag of different low-energy ranges (blue) compared to the lightcurve in the 150--350 keV range as well as the entire 8--100 keV range (orange). {\it [Right]} The CCF coefficient as a function of the low-energy lightcurve (60--100 keV) lag. The red band shows the estimation of the trend.}
	\label{fig:Spectral_Lags}
\end{figure}

Owing to the paucity of data above $\sim$300 keV, we constrain our inspection of spectral lag to energies below 300 keV.  First, we broadly compare the lightcurve in 8--100 keV to the lightcurve in 150--300 keV.  We compute the lag using 64 ms binned data for the lightcurve ranging from T0-0.32 s to T0+0.768 s and find a slight preference for a soft lag of $+150^{+106}_{-140}$ ms. We also sub-divide the low-energy interval into five energy ranges, and calculate the lag in each of those sub-ranges relative to the 150--300 keV lightcurve.  As shown in Figure~\ref{fig:Spectral_Lags}, we do not find any significant evolution of spectral lag as a function of energy. There is a preference for a soft lag of $\sim$100 ms; however, due to large uncertainties, this is still generally consistent with zero.  We also show in Figure~\ref{fig:Spectral_Lags} the DCCF as a function of time lag for the best constrained low-energy interval: 60--100 keV relative to 150--300 keV.

\subsection{Minimum Variability Timescale}\label{sec:varibility}
The minimum timescale on which a GRB exhibits significant flux variations has long been thought to provide an upper limit as to the size of the emitting region and yield clues to the nature of the burst progenitor\mod{~\citep{Schmidt78,Fenimore93}}. Here we employ a structure function (SF) estimator, based on non-decimated Haar wavelets, in order to infer the shortest timescale at which a GRB exhibits uncorrelated temporal variability.  This technique was first employed in \citet{Kocevski2007} to study the variability of X-ray flares observed in afterglow emission associated with Swift detected GRBs, and further developed by \citet{Golkhou2014} and \citet{Golkhou2015} for use in \Swift BAT and GBM data, respectively.  Here we follow the method outlined in~\citet{Golkhou2015} in applying the SF estimator to GBM TTE data.  We summed 200 $\mu$s resolution light curve data for NaIs 1, 2, and 5 over an energy range of 10--1000 keV.  We subtracted a linear background model estimated from T0 $\pm10$ s, which excludes data from the \Tninety interval.  

\begin{figure}
	\subfigure[]{\label{fig:mvt}\includegraphics[scale=0.45]{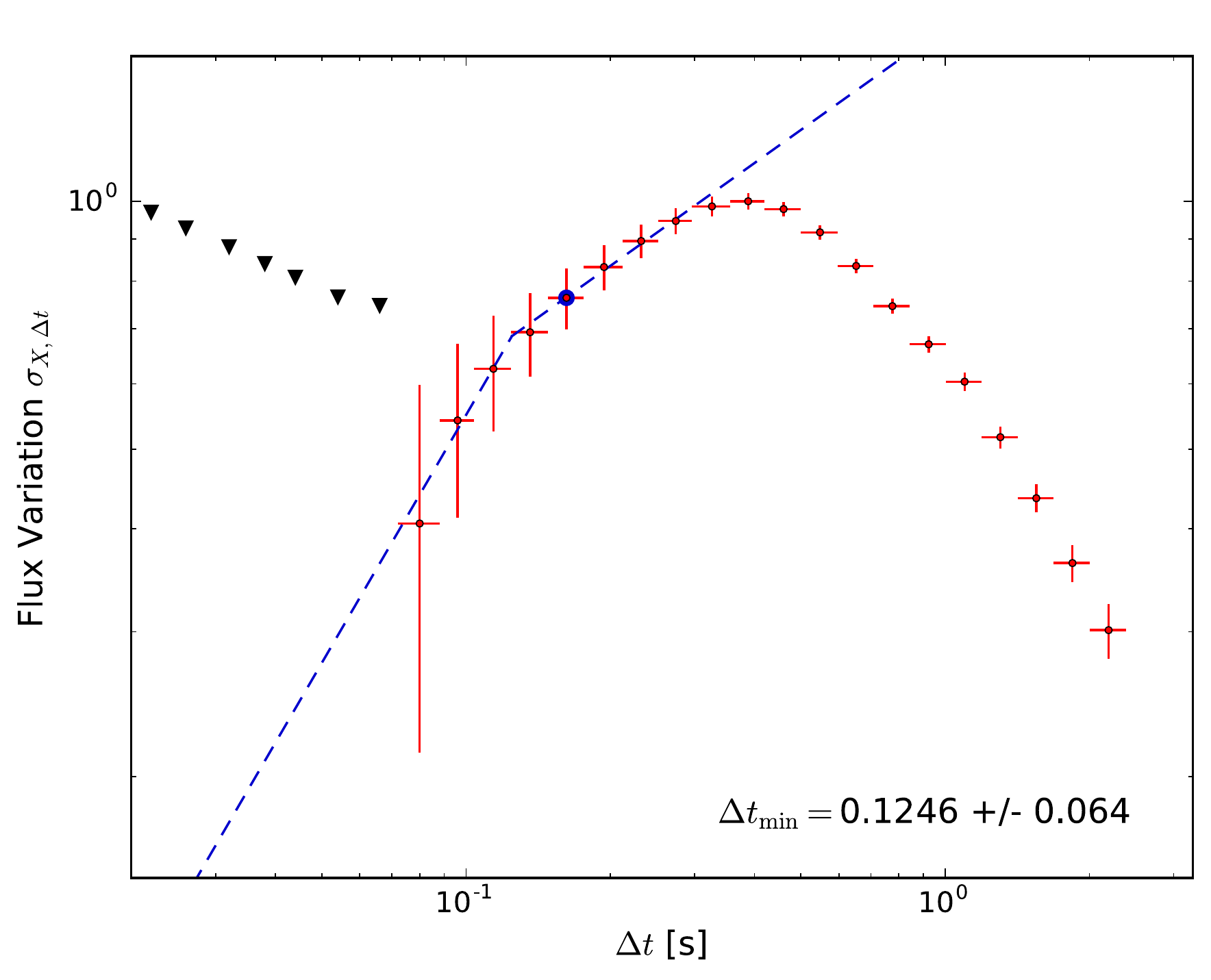}}
	\subfigure[]{\label{fig:golkhou}\includegraphics[scale=0.55 ]{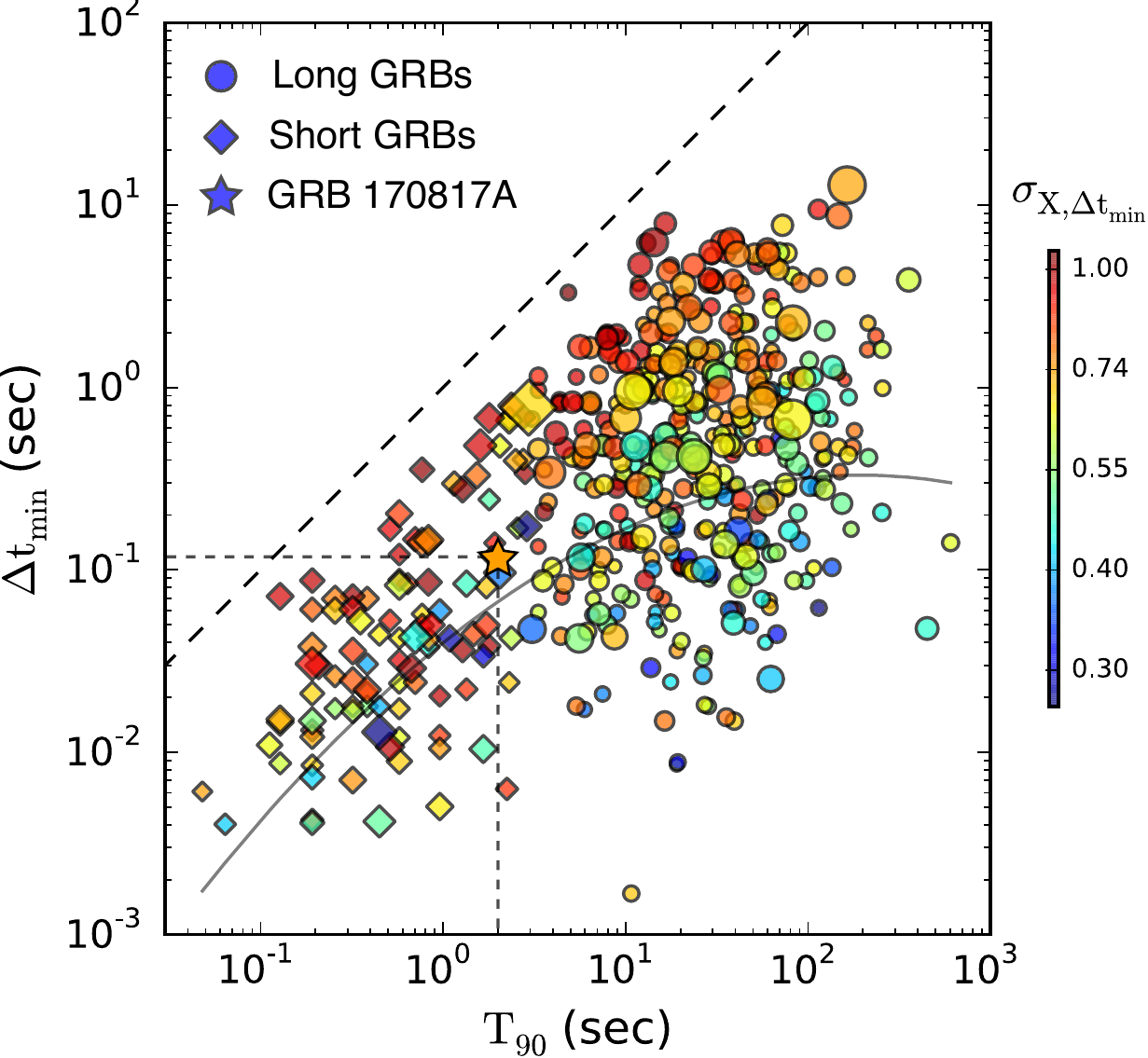}}
	\caption{{\it [Left]} Haar wavelet scaleogram vs. $\Delta$t for GRB~170817A.  The red points represent 3$\sigma$ excesses over the power associated with Poisson fluctuations at a particular timescale and the triangles denote $3\sigma$ upper limits. The break in the blue dashed line represents the shift between correlated and uncorrelated variability in the data, which we define as the minimum variability timescale. {\it [Right]} Comparison of $\Delta t_{\rm min}$ for GRB~170817A to other GBM-triggered GRBs analyzed by \citet{Golkhou2015}. The colors represent the flux variation level at $\Delta t_{\rm min}$.  \mod{The symbol sizes are proportional to the ratio between the minimum variability and S/N timescale, defined as the shortest observable timescale with power above the Poisson noise floor.} The solid curved line is the typical S/N timescale as a function of T90 and the dashed line shows the equality line. Figure reproduced from \citet{Golkhou2015}. \label{fig:Variability}}
\end{figure}

The resulting Haar scaleogram showing the flux variation level (i.e.\ power) as a function of timescale can be seen in Figure~\ref{fig:mvt}. The red points represent 3$\sigma$ excesses over the power associated with Poisson noise at a particular timescale and the triangles denote $3\sigma$ upper limits. We define the minimum variability timescale as the transition between correlated (e.g. smooth, continuous emission) and uncorrelated (e.g. rapid variations or pulsed emission) variability in the data. \mod{As discussed in \citet{Golkhou2014}, the resulting minimum variability timescale $\Delta t_{\rm min}$ does not necessarily represent the shortest observable timescale in the light curve, which tends to heavily depend on the signal-to-noise of the data.  Rather it is a characteristic timescale that more closely resembles the rise time of the shortest pulses in the data.} Such correlated variability appears in the scaleogram as a linear rise relative to the Poisson noise floor at the smallest timescales and the break in this slope represents the shift to uncorrelated variability. The linear rise phase and the subsequent break are demarcated by the dashed blue line. The blue circle marks the extracted value of $\Delta t_{\rm min}$.  

Using the full 10--1000 keV energy range, we obtain $\Delta t_{\rm min}=0.125 \pm 0.064$ s. \mod{Repeating the analysis over two restricted energy ranges covering 10--50 keV and 10--300 keV, we obtained values of 0.312 $\pm$ 0.065 s and 0.373 $\pm$ 0.069 s respectively. A decrease in $\Delta t_{\rm min}$ as a function of increasing energy matches the results reported by \citet{Golkhou2015} and is consistent with the observed trend of GRB pulse durations decreasing as a function of energy, with hardest energy channels having the shortest observed durations~\citep{Fenimore1995, Norris1996, KocevskiLiangLags}. Figure~\ref{fig:golkhou} shows the resulting $\Delta t_{\rm min}$ over 10--1000 keV energy range compared to the full sample of short and long GBM-detected GRBs analyzed by \citet{Golkhou2015}. It is apparent that GRB~170817A is broadly consistent with the short GRB population.}

\subsection{Search for Periodic Activity}\label{sec:period}
Some short GRB models invoke a newly born millisecond magnetar as a central engine, e.g., \cite{Bernardini2015}. The GBM TTE data were searched for evidence of periodic activity during or immediately before and after the burst that might indicate the pulse period of the magnetar. For two energy ranges, 8--300 keV and 50--300 keV, three time intervals were searched: T0-10 s to T0+10 s, T0-2 s to T0+2 s, and T0-0.4 s to T0+2.0 s, selected by eye to incorporate all possible emission from the burst. The TTE data were binned into 0.25 ms bins and input into PRESTO\footnote{\url{http://www.cv.nrao.edu/\%7Esransom/presto/}} \citep{Ransom2001}, a standard software suite used for searches for millisecond pulsars in \Fermi/LAT, X-ray, and radio data. Specifically, an accelerated search \citep{Ransom2002} was used to search for drifting periodic signals in the range 8-1999 Hz. Significant red noise, due to the variability of the burst itself was found at lower frequencies. No significant periodic signals were detected above $1.5\sigma$ that were present in all energy ranges and time intervals. To search for quasi-periodic signals, each time interval above was divided into subintervals (1 s, 0.5 s, and 0.4 s, respectively). Power spectra were generated for each sub-interval and then were averaged over each full time range. No significant quasiperiodic signals were found in any of the time intervals in either energy range. Red noise was present below about 1-2 Hz, consistent with the the noise in the periodic searches. \mod{The power above 1-2 Hz was consistent with white noise.}

\subsection{Pulse Shape and Start Time}\label{sec:PulseShape}
GRB pulse shapes can be well described by analytic functions~\citep{Norris+96pulse,Norris+05pulse,Bhat+12pulse}. These are especially useful to derive more accurate estimates of pulse properties in case the GRB is dim. We adapt the pulse profile described in \citet{Norris+96pulse}, where the pulse shape is given by $I(t)=A \exp{(-((t_{\rm peak}-t)/\sigma_{\rm rise})^\nu)}$ for $t<t_{\rm peak}$ and $I(t)=A \exp{(-((t-t_{\rm peak})/\sigma_{\rm decay})^\nu)}$ for $t>t_{\rm peak}$. Here $A$ is the amplitude at the peak time of the pulse, $t_{\rm peak}$, $\sigma_{\rm rise}$ and $\sigma_{\rm decay}$ are the characteristic rise and decay times of the pulse respectively. \mod{\citet{Lee+00peakedness} studied a large sample of GRB pulses, including short GRBs by fitting the same function. While they did not discuss short GRBs in particular, $\nu=2$ is close to the median of the distribution of fitted $\nu$ values.  We therefore fix the shape parameter to $\nu=2$ which also aids the convergence of the fit.}

By fitting the summed lightcurve of NaI 1, 2, and 5 with 32 ms resolution we find the shape of the main pulse is described by $t_{\rm peak}=-114 \pm 45$ ms, $\sigma_{\rm rise} =129 \pm 54$ ms, and $\sigma_{\rm decay} = 306 \pm 64$ ms \mod{($\chi^2_r=0.99$ for 276 degrees of freedom)}. We define the start time as the time where the pulse reaches the 10\% of its peak and from these pulse parameter values we find  $t_{\rm start}=-310 \pm 48$ ms, relative to T0.  This pulse shape is used in Section~\ref{sec:Detectability} for the production of a synthetic GRB to estimate its detectability at weaker than observed intensities. 

\section{Limits on Other Gamma-ray Emission}\label{sec:limits}
Aside from the prompt emission that triggered GBM, we investigate other possible associated gamma-ray signals: precursors and extended emission lasting several seconds. \mod{While precursors and extended emission have been observed for some short GRBs, claims of long-term or flaring emission on the timescale of several hours or days is rarer. Due to the proximity of the GRB within the GW observing horizon, we also search for persistent emission from GRB~170817A at hard X-ray energies that might, for example, be associated with afterglow emission from the source (e.g.,~\citet{nustar130427a}).}

\subsection{Limits on Precursors\label{sec:precursors}}
Evidence for precursor emission has been found for GRBs detected by \Swift BAT \citep{0004-637X-723-2-1711}, and their existence has been searched for in the SPectrometer for INTEGRAL AntiCoincidence Shield \citep{2016arXiv161202418M}.  There are several theoretical models to explain their existence (e.g. \citealt{2012PhRvL.108a1102T}, \citealt{MetzgerPrecursors}), and some ideas have been developed for their use in joint GW-EM detections (e.g. \citealt{schnittman2017electromagnetic}). In the GW era, there has been an increase in interest in precursors, as this emission is less relativistically beamed, or potentially isotropic, and might be observable to larger inclination angles than the prompt short GRB emission. A search for precursor emission associated with GBM detected GRBs with \Tninety $< 2$ s was performed by \citet{burns2017searching}. This work was intended to inform the time range expected for EM counterparts to GWs, \mod{but little evidence was found for precursor activity in the GBM archive of more than 300 short GRBs, with few exceptions~\citep[e.g.][]{grb090510}.}

\mod{The largest time offset claimed for possible precursor emission before a short GRB is $\sim$T0-140 s \citep{0004-637X-723-2-1711}, therefore} we use the targeted search~\citep{Blackburn15,Goldstein16} to examine an interval covering T0-200 s to T0, which encompasses all reported putative short GRB precursors offsets and most expected offsets from theoretical and numerical modeling. We note that the lowest \Epeak among the spectral templates used in the targeted search is only 70 keV, so this search is not especially sensitive to weak events with peak energies below a few tens of keV. We find no significant emission before T0.


Therefore, with no detected precursor signals, we calculate upper limits on precursor emission for GRB~170817A using the procedure described in~\citet{FermiGW151226} and \citet{FermiGW170104}. We set a range of upper limits based on three template spectra which we use in our targeted search, generally referred to as a `soft,' `normal,' and `hard' template.
\mod{Using these templates and assuming a 0.1 s (1 s) duration precursor up to 200 s before the GRB, we find a $3\sigma$ flux upper limit range of $6.8-7.3\times10^{-7}$ ($2.0-2.1\times10^{-7}$) $\rm erg \ s^{-1} \ cm^{-2}$ for the soft template, $1.3-1.5\times10^{-6}$ ($3.9-4.2\times10^{-7}$)  $\rm erg \ s^{-1} \ cm^{-2}$ for the normal template, and $3.4-3.7\times10^{-6}$ ($9.8-11\times10^{-7}$) $\rm erg \ s^{-1} \ cm^{-2}$ for the hard template.}

\subsection{Limits on Extended Emission}
Since the launch of \Swift BAT, a class of short GRBs with softer extended emission has been discovered \citep{norris2006short}, with a signature short, hard spike typical of short GRBs followed by a weaker long, soft tail extending from a few seconds to more than a hundred seconds. For a source position within its coded-mask field-of-view, \Swift BAT background rates are low and relatively stable compared to the high and variable background flux experienced by the uncollimated GBM detector. Therefore, GBM typically does not find evidence of the extended emission unless a GRB is very bright, in which case the extended emission can contribute tens of seconds to the GBM-estimated \Tninety.  In general, however, \Swift BAT detects extended emission when GBM does not, although mild evidence for that emission can be found in the GBM data starting with the knowledge from the BAT data that it exists.

We find no evidence for extended emission, though we note that the sensitivity to such extended emission was not optimal owing to the proximity of \Fermi to the SAA and the resulting higher and more variable background rates than elsewhere in the \Fermi orbit. Using the same procedure as in Section~\ref{sec:precursors}, \mod{we estimate a $3\sigma$ flux upper limit range averaged over a 10 s of $6.4-6.6\times10^{-8} \ \rm erg \ s^{-1} \ cm^{-2}$ for the presence of any soft extended emission out to 100 s after the GBM trigger.}

\subsection{Long-Term Gamma-ray Emission Upper Limits}
To estimate the amount of persistent emission during a 48 hour period centered at T0, we use the Earth Occultation technique \citep{2012ApJS..201...33W} to place $3\sigma$ day-averaged flux upper limits over the 90\% credible region of the HLV skymap.  We use a coarse binning resolution on the sky to inspect the contribution of persistent emission over the sky map, and we compute a flux upper limit for each bin that has been occulted by the Earth at least 6 times in both the 24 hour period preceding and following T0. This is done to ensure some minimal statistics to compute a day-averaged flux. Nearby known bright sources, determined from \Swift BAT monitoring, are automatically included in the model fit and thus accounted for in any calculations of flux from the GRB source. In addition, position bins that contain known bright, flaring sources are removed during post-filtering of the data.  The range and median of the flux upper limits over the sky map are shown in Table~\ref{table:upper_limits}, and are consistent with the observed background \mod{on this timescale. Therefore, we find no evidence for persistent emission from GRB~170817A, which is typical for GBM observations of GRBs.}

\section{Detectability of GRB~170817A}\label{sec:Detectability}
GBM triggered on this GRB despite an increasing background as \Fermi approached the SAA, primarily because the source--detector geometry was near optimal for the triggering detectors. 
After the GBM flight software triggers, it continues to evaluate the remaining trigger algorithms. Three other trigger algorithms also exceeded their threshold: a 256 ms interval ending 128 ms after the trigger time, and 512 ms and 1024 ms intervals that ended 256 ms after the trigger time.  The significances were $5.16\sigma$,  $6.25\sigma$ and $4.52\sigma$, respectively. All four of the trigger algorithms that exceeded their thresholds for this GRB were for the energy range 50--300 keV and are GBM trigger algorithms that typically detect short GRBs. As the thresholds for all four algorithms are $4.5\sigma$, the most sensitive algorithm was the one based on the 512 ms accumulation.  At $6.25\sigma$, this GRB could have been dimmed to $\sim 70$\% of its observed brightness and it would still have triggered GBM.  The precise sensitivity to a similar GRB at a different time depends on the direction of the GRB relative to \Fermi, the background in the GBM detectors at the time of the GRB, and the phasing of the accumulations used for triggering relative to the GRB pulse profile.

If the flight software were unable to trigger on a weaker version of this GRB, the offline searches developed for multi-messenger counterpart searches of GBM data have lower detection thresholds.  The targeted search found this GRB with the `normal' spectral template with a search SNR of 12.7. If we set the reporting threshold for an event of interest at a false alarm rate of $\sim 10^{-4}$ Hz (SNR $\sim5.4$), then this GRB could be weakened to $\sim43$\% of its observed brightness and still be detected with the targeted search, assuming the same background and detector--source geometry. 

Similarly, the  untargeted search for short GRBs could detect this event with medium confidence if it were weakened to $\sim50$\% of its observed brightness estimated from simulations of the efficiency of the search using a synthetic pulse injected into suitable background data~\citep{untargetedsearch}. Background data from 30 orbits after T0 was used, when \Fermi was located in a similar position in its orbit and was in the same orientation.  Simulated GRBs were added to the data, using the pulse shape found in section \ref{sec:PulseShape} and a cutoff power-law spectrum from the fit described in section \ref{sec:spectra}.  The intensity was reduced until the simulated GRB was just found at a quality that would result in a \Fermi-GBM Subthreshold GCN Notice with a medium reliability score.

\section{Summary}
We presented observations by \Fermi GBM of the first GRB associated with a binary neutron star merger. Our observations show GRB~170817A to most likely be a short--hard GRB, although it appears softer than the typical short GRB detected by GBM.  The progenitors of short--hard GRBs have been hypothesized to be mergers of compact binary systems, at least one member of which is a neutron star, which is directly confirmed for GRB~170817A by the associated GW emission~\citep{capstonepaper,jointpaper}. Comparing the standard analysis results to the GBM GRB catalog, we find that this GRB has a lower peak energy than the average short GRB, but may exhibit a harder power-law index. In terms of the 64 ms peak flux, it is one of the weakest short GRBs that GBM has triggered on, though owing to its $\sim 2$ s duration, it has near-median observed fluence.

\mod{A more detailed analysis of this GRB uncovers some interesting results.  The basic properties (peak energy, spectral slope, and duration) of the main peak are broadly consistent with the leading prompt emission models (e.g.\ dissipative photosphere \citep{Rees+05photdis} or internal shocks \citep{Rees+94is}). Aside from the main peak of $\sim0.5$ s in duration, there appears to be softer emission lasting for $\sim1.1$ s, which has a localization that is in agreement with both the localization of the main peak and the HLV sky map. This emission strongly favors a blackbody spectrum over the typical power-law found when fitting background-subtracted noise in GBM. If this weak soft emission is associated with GRB~170817A, there are some interesting implications, although the fact that the $\sim 1$ s long potentially thermal soft emission is after the hard non-thermal emission makes it difficult to interpret as typical GRB photospheric emission. There is evidence for a thermal component in some GRBs, but it occurs at the same time as the dominating non-thermal emission~\citep[e.g.][]{Zhang+09mag, Guiriec+13sgrbbb}, or the dominant emission itself can be modeled as a quasi-thermal spectrum which broadens in time to approximate a non-thermal spectrum, possibly a sign of photospheric dissipation~\citep{ryde090902b}.  Significantly weaker and lower temperature thermal emission has been observed \citep{Starling+12bbxrt} a few hundred seconds after trigger with \Swift XRT, albeit for long GRBs.  As discussed in Section~\ref{sec:spectra}, one potential explanation for the presence of this emission is photospheric emission from a cocoon, which is thought to be visible from larger viewing angles than the typical uniform-density annular GRB jet~\citep{Lazzati+17cocoon}.} 

Aside from this intriguing soft emission, we find no evidence for precursor emission, several second-long extended emission, or day-long flaring or fading from the source.  We have also calculated the spectral lag, which we find is consistent with zero at $\sim 1 \sigma$, primarily owing to the weakness of the GRB in the GBM data. However, we do find a systematic preference for a positive (soft) lag, which may be an indication for hard-to-soft spectral evolution within the GRB main pulse.  A calculation of the minimum variability timescale for GRB~170817A shows that it is consistent with the shorter variability timescales observed in short GRBs.  A search for periodic emission associated with the GRB did not find any significant periodic or quasi-periodic activity preceding or following the GRB.

\Fermi GBM, with instantaneous coverage of 2/3 of the sky and with high uptime ($\sim85$\%), is a key instrument for providing EM context observations to gravitational observations.  Joint GW-EM detections with GBM can allow for the confirmation of progenitor types of GRBs, set constraints on fundamental physics such as the speed of gravity and Lorentz Invariance Violation, and further constrain the rates of multimessenger sources.  In a companion paper~\citep{jointpaper} some of these analyses are performed for GRB~170817A. Future GBM triggers may, with latencies of ten seconds to minutes, provide localizations that can reduce the joint localization region, particularly when only a subset of GW detectors are online at the time of a detection. The offline searches can provide localizations for weaker events at delays of up to a few hours.\\

\noindent
\mod{We dedicate this paper to the memory of Neil Gehrels who was an early and fervent advocate of multi-messenger time-domain astronomy and with whom we wish we could have shared the excitement of this tremendous observation.}\\

\noindent
\mod{We thank the referee for an exceptionally prompt and helpful report.} The USRA co-authors gratefully acknowledge NASA funding through contract NNM13AA43C. The UAH co-authors gratefully acknowledge NASA funding from co-operative agreement NNM11AA01A. E.B. and T.D.C. are supported by an appointment to the NASA Postdoctoral Program at the Goddard Space Flight Center, administered by Universities Space Research Association under contract with NASA. D.K., C.A.W.H., C.M.H., and T.L. gratefully acknowledge NASA funding through the \Fermi GBM project. Support for the German contribution to GBM was provided by the Bundesministerium f{\"u}r Bildung und Forschung (BMBF) via the Deutsches Zentrum f{\"u}r Luft und Raumfahrt (DLR) under contract number 50 QV 0301. A.v.K. was supported by the Bundesministeriums für Wirtschaft und Technologie (BMWi) through DLR grant 50 OG 1101. N.C. and J.B. acknowledge support from NSF under grant PHY-1505373. S.M.B acknowledges support from Science Foundation Ireland under grant 12/IP/1288.  

\bibliographystyle{aasjournal}
\bibliography{references}

\clearpage

\clearpage

\linespread{1.0}

\begin{table}
\begin{singlespace}
\caption{The GBM timeline of the trigger and reporting of GRB~170817A \label{table:GBM_timeline}}
\end{singlespace}
\begin{tabular}{|l| c | l |}
\hline
Time (UTC)	&	Relative	&	Comment	\\
\hline
\hline
12:41:06.474598  &   0    &  Trigger Time: \\
 & & End of 0.256 s interval containing statistically significant rate increase \\
\hline
12:41:06.477006  &   +2.4 ms  & Triggered: \\
 & & Autonomously detected in-orbit by the \Fermi GBM flight software \\
\hline
12:41:20      &     +14 s     &  \Fermi GBM Alert Notice sent by the GCN system at NASA/GSFC \\
\hline
12:41:31     &      +25 s    &   Automatic location from GBM flight software sent by the GCN: \\
 & & RA=172.0, Dec=-34.8, err=32.6 deg \\
\hline
12:41:44     &      +38 s   &    More accurate automatic location by ground software sent by GCN: \\
& & RA=186.6, Dec=-48.8, err=17.4 deg \\
\hline
13:26:36    &     +44.9 min    &   More accurate human-guided localization sent by GCN: \\
 & & RA=176.8, Dec=-39.8, err=11.6 deg \\
 \hline
13:47:37    &     +66.5 min    &   LVC GCN Circular reporting localization and consistency of signal \\
& & with a weak short GRB~\citep{gcnGBMGRB170817A_1} \\
 \hline
20:00:07    &     +7.3 hr    &   Public GCN Circular establishing GRB name and \\
& & standard GBM analysis~\citep{gcn170817Apublic}\\
\hline
00:36:12 &     +11.9 hr    &   LVC GCN Circular reporting updated spectral analysis, \\
(next day) &  & energetics, and association significance~\citep{gcnGBMGRB170817A_2}\\
\hline
\end{tabular}
\end{table}

\begin{table}
\begin{singlespace}
\caption{The angle from each GBM detector normal to the OT position.  The highlighted detectors were used for analysis. \label{table:detector_angles}}
\end{singlespace}
\begin{tabular}{lcl}
\hline
Detector	&	Angle ($^\circ$)	&	Comment	\\
\hline
NaI 0 & 63 & \\
\rowcolor{gainsboro} NaI 1 & 39 & Good geometry \\
\rowcolor{gainsboro} NaI 2 & 15 & Good geometry \\
NaI 3 & 86 & \\
NaI 4 & 101 & \\
\rowcolor{gainsboro} NaI 5 & 42 & Good geometry \\
NaI 6 & 104 & blocked by spacecraft \\
NaI 7 & 130 & blocked by spacecraft \\
NaI 8 & 167 & blocked by spacecraft \\
NaI 9 & 86 & blocked by LAT radiator \\
NaI 10 & 78 & blocked by LAT radiator \\
NaI 11 & 138 & blocked by LAT radiator \\
\hline
\rowcolor{gainsboro} BGO 0 & 44 & Good geometry \\
BGO 1 & 136 & blocked \\
\hline
\end{tabular}
\end{table}

\begin{table}
\centering
 \begin{tabular}{ccccccc}
 Time Range (s) &    Model    &    \Epeak (keV)    &    Index    &    kT (keV)    & \begin{tabular}{@{}c@{}}Energy Flux\\ (10$^{-7}$ erg s$^{-1}$ cm$^{-2}$)\end{tabular} & 
\begin{tabular}{@{}c@{}}Energy Fluence\\ (10$^{-7}$ erg cm$^{-2}$)\end{tabular}\\ 
\hline
\multicolumn{7}{ c }{Standard Analysis} \\ \hline
$-$0.192:0.064 & Comptonized & 215 $\pm$ 54 & 0.14 $\pm$ 0.59  & - & 5.5 $\pm$ 1.2 & 1.4 $\pm$ 0.3  \\
$-$0.128:-0.064 & Comptonized & 229 $\pm$ 78 & 0.85 $\pm$ 1.38  & - & 7.3 $\pm$ 2.5 & 0.5 $\pm$ 0.2 \\
\hline
\multicolumn{7}{ c }{Detailed Analysis} \\ \hline
$-$0.320:0.256 & Comptonized & 185 $\pm$ 62 & $-$0.62 $\pm$ 0.40  & - & 3.1 $\pm$ 0.7 & 1.8 $\pm$ 0.4  \\
0.832:1.984 & Blackbody & - & -  & 10.3$\pm$1.5 & 0.53 $\pm$ 0.10 & 0.61 $\pm$ 0.12 \\
 \end{tabular}
 \caption{The spectral fit parameters for the standard GBM analysis and the detailed analysis of the main pulse and softer emission. Note the time range is relative to the GBM trigger time.  These spectral fits are considered the best fits to the data during the corresponding times.\label{tab:Spectra}}
 \end{table}

{\tiny
\begin{table}
\centering
 \begin{tabular}{c | ccccc}
Energy (keV)    &  &  Min    &    Max    &    Median    &  \\ 
\hline
12-27 & &0.84 & 2.06 & 1.31 &\\
27-50 & &0.93 & 2.28 & 1.42 & \\
50-100 & &1.58 & 3.95 & 2.37 &\\
100-300 & &3.34 & 8.73 & 5.14 & \\
300-500 & &7.29 & 20.6 & 11.4 & \\
\hline
12-100 & &1.45 & 1.80 & 1.59 &\\
 \end{tabular}
 \caption{$3\sigma$ 24-hour flux upper limits (units of $10^{-9}$ erg s$^{-1}$ cm$^{-2}$ ) over the HLV map.\label{table:upper_limits}}
 \end{table}
}
%
\end{document}